\documentclass{article}
\usepackage{amsfonts,amsmath,amssymb,authblk,floatrow,fullpage,graphicx,tabularx,threeparttable,tikz,pgfplots,xcolor,lscape,float}
\usepackage[colorlinks]{hyperref}
\usepackage[noabbrev,capitalise,sort&compress]{cleveref}
\usepackage{subcaption}
\usepackage{appendix}
\usepackage[textsize = small]{todonotes}
\usepackage[nohyperlinks,nolist]{acronym}
\usetikzlibrary{arrows,arrows.meta,calc,fit}

\newcommand{\dd}{\mathrm{d}}                    
\renewcommand{\vec}[1]{\boldsymbol{#1}}         
\newcommand{\mat}[1]{\boldsymbol{#1}}           

\newcommand{\pd}{\partial}                      
\newcommand{\parder}[2]{\frac{\pd #1}{\pd #2}} 

\DeclareMathOperator*{\Aop}{\mathbb{A}}

\newcolumntype{P}[1]{>{\centering\arraybackslash}p{#1}}
\newfloatcommand{capbtabbox}{table}[][\FBwidth]

\definecolor{ccao}{rgb}{0.596, 0.251, 0.043}
\definecolor{ccaa}{rgb}{0.314, 0.882, 0.957}
\definecolor{ael}{rgb}{0.063, 0.714, 0.188}
\definecolor{elel}{rgb}{0.478, 0.953. 0.373}
\definecolor{elc}{rgb}{0.584, 0.106, 0.478}
\definecolor{cccc}{rgb}{0.796, 0.145, 0.129}
\definecolor{elccc}{rgb}{0.216, 0.114, 0.781}
\definecolor{ccco}{rgb}{0.533, 0.322, 0.055}

\begin{acronym}
    \acro{scl}[SCL]{space-charge layer}
\end{acronym}
\pgfplotsset{compat=newest}
\begin{document}
\title{A Finite Element Formulation to Three-Dimensionally Resolve Space-Charge Layers in Solid Electrolytes}
\author{Stephan Sinzig$^{1,2}$, Thomas Hollweck$^{1}$, Christoph P. Schmidt$^{1}$, Wolfgang A. Wall$^{1}$}
\affil{\small $^1$ Institute for Computational Mechanics, Technical University of Munich, Boltzmannstra\ss e 15, 85748 Garching bei M\"unchen, Germany \\ $^2$ TUMint.Energy Research GmbH, Lichtenbergstra\ss e 4, 85748 Garching bei M\"unchen, Germany}
\date{}
\maketitle
\section*{Abstract}
\label{sec:abstract}
All-solid-state batteries are seen as promising candidates to replace conventional batteries with liquid electrolytes in many applications. However, they are not yet feasible for many relevant applications. One particular question of interest is the identification of physical effects inside all-solid-state batteries and their quantitative influence on the performance of the entire battery cell. Simulation models can contribute to answering the aforementioned question by systematical studies, e.g. enabling or disabling certain physical effects. Especially the influence of space-charge layers (SCLs) is heavily discussed in the scientific community. So far, the different length scales of SCLs and the microstructure of a battery cell made a spatial discretization of realistic microstructures with resolved SCLs infeasible. However, thermodynamically consistent continuum models which are applied to simplified geometries are already established in the literature. In this work, we propose a model that enables the prediction of the spatial development of SCLs within geometrically resolved microstructures by exploiting that effects in SCLs are predominantly one-dimensional. With the proposed approach it is possible to quantify the geometric influence of realistic microstructures on the formation process of SCLs. SCLs in realistic microstructures remarkably differ from SCLs computed with simplified one-dimensional models which are already established in the literature.
\section*{Introduction}
\label{sec:intro}
Research activity has strongly increased in recent years to improve both the energy and power densities of batteries. Especially, lithium-ion batteries are nowadays seen as the superior battery technology for many applications~\cite{Janek2016}, especially for electric vehicles. It is foreseeable, that conventional lithium-ion batteries with liquid electrolytes will reach their physical limit soon. All-solid-state batteries could theoretically overcome the drawbacks of conventional lithium-ion batteries with liquid electrolytes. Some of the advantages of all-solid-state batteries are their possibility for high power densities, facilitating of lithium metal anodes, thus achieving high energy densities, and ensuring high safety standards due to the non-flammability of many solid electrolytes~\cite{Takada2013, Fu2022, Famprikis2019}. However, they still require more research effort to establish them for various real-world applications. While experimental research is already well-established in the field of electrochemistry, simulative investigations become more prominent to rapidly evaluate the influence of different operating scenarios, quantifying the influence of certain physical effects, or testing the combination of different materials. Profound predictions of the behavior of a battery cell require models that are based on fundamental physics and are solved in a mathematically consistent manner.\\
The formation of regions where charges separate, i.e.~double layers for liquid electrolytes~\cite{Stern1924} and \acp{scl} for solid electrolytes~\cite{Lehovec1953}, is known for decades. However, their influence in terms of resistance and capacitance on the entire battery cell is heavily discussed in the literature~\cite{Klerk2018}, ranging from rather negligible~\cite{Haruta2015, Tateyama2019} to important~\cite{Yu2017a, Haruyama2014, Luntz2015}. We do not claim to give a full overview of the physical phenomena inside \acp{scl} but refer the reader to the literature, e.g.~\cite{Usiskin2020}, and summarize only the key aspects. \acp{scl} are small regions inside the solid electrolyte, that form close to the electrodes. Inside these regions, separation of charges is observable (e.g.~\cite{Newman2004, Latz2011}), as shown in various experiments~\cite{Cheng2020, Katzenmeier2021a, Katzenmeier2021, Katzenmeier2022, Wang2020b, Nomura2019, Liu2022}. Due to the different chemical potentials of two materials in contact, charge carriers will redistribute to form either an accumulation layer or a depletion layer until equilibrium is reached~\cite{Zhang2022}. \acp{scl} occur at all interfaces of a battery cell where two materials with different chemical potentials are in contact. They occur especially at the interface between the solid electrolyte and the electrodes, and at internal interfaces inside the solid electrolyte~\cite{Chen2016}.\\
Simulation models can contribute to the mentioned discussion of the influence of \acp{scl} on the entire battery cell. Of course, a simulation model can never cover all physical effects, that occur inside an all-solid-state battery but needs to be tailored to the specific question that it should answer. In the scientific community, different modeling approaches are available to incorporate the effect of \acp{scl} into a model, each with a different focus: Atomistic models (e.g.~\cite{Fingerle2017, Vatamanu2017}), DFT models (e.g.~\cite{Haruyama2014, Stegmaier2017}), kinetic Monte Carlo models (e.g.~\cite{Katzenmeier2022a}), or continuum models. In light of the following sections, we want to elaborate more on continuum models, as the other models can only be applied to domains with dimensions in the range of nanometers due to computational limitations and thus, not to geometries representing realistic microstructures. Continuum models can be further subdivided into: Phenomenological models, which modify known equations e.g.~from liquid electrolytes to include the effects of solid electrolytes (e.g.~\cite{Swift2019, Swift2021, Brogioli2019}), zero-dimensional models, that resolve a complex geometric microstructure and add e.g. a capacitor to the interface to represent \acp{scl} (e.g.~\cite{Lueck2018, Hein2020, Lueck2019}), and one-dimensional models, that spatially resolve the shape of \acp{scl} between two electrodes (e.g.~\cite{Braun2015, BeckerSteinberger2021}). As all models, the outlined models have different limitations that can be significant for relevant questions. Phenomenological models do not ensure positive entropy production and are thus thermodynamically not consistent or neglect the transient development of the \ac{scl}, zero-dimensional models cannot resolve the spatial shape of \acp{scl}, and one-dimensional models neglect the inhomogeneous geometric influence, which can be significant as we will show in this work.\\
Especially the different length scales ($\approx 10 - 100 \ \text{nm}$~\cite{Katzenmeier2021}) of \acp{scl} and realistic microstructures ($\approx 100 \ \mu \text{m}$) are currently hindering a three-dimensionally resolved solution of the \acp{scl} in realistic microstructures. A three-dimensional mesh, as needed for the discretization of the continuous model, e.g. with the finite element method, would require mesh cells with a size of about~1~nm to capture gradients within the thin layer. This would exceed currently available computational resources if realistic microstructures would be discretized with the required fineness, resulting in the order of one billion nodes of a discretization mesh.\\
In this work, we introduce a novel approach that allows resolving and incorporating \acp{scl} within geometrically complex microstructures. The approach is motivated by the observation, that \acp{scl} develop in regions close to the electrodes and are predominantly one-dimensional due to the perpendicular electric field on equipotential surfaces, i.e.~perfectly electronic conducting electrodes. In the remaining part of the solid electrolyte, the condition of local charge neutrality holds~\cite{Latz2011}. By using different discretizations in \ac{scl} regions and outside of \ac{scl} regions we propose a solution to the unsolved challenge to resolve \acp{scl} in realistic microstructures. Based on the observation of predominant one-dimensional effects in \acp{scl} we discretize the domain in the vicinity of the electrodes in one dimension, while it is three-dimensionally discretized outside of the \ac{scl} region. This reduces the computational effort significantly because the mesh size outside of the \ac{scl} region can be adapted according to the dimensions of the microstructure, while the one-dimensional discretization in the \ac{scl} region can be adapted to the shape of the \ac{scl}. Consequently, the computational effort reduces to a manageable size. We base the continuous \ac{scl} model on the work reported in~\cite{Braun2015} and~\cite{BeckerSteinberger2021}, which guarantees positive production of entropy and is formulated in the three-dimensional space. However, our proposed discretization scheme is not attached to this model but is conceptually applicable to any other continuum model for \acp{scl}. Additionally, we reduce the computational effort, by enforcing the condition of constant concentrations to regions outside of the \ac{scl}.\\
This work is outlined as follows: We begin with recalling a continuous approach to model \acp{scl} including physically meaningful boundary and initial conditions. Afterwards, we simplify the model outside of the \ac{scl} regions by enforcing constant concentrations. Based on this, we introduce a novel approach for a consistent coupling of \ac{scl} regions and regions outside of the \ac{scl}. Subsequently, we present the numerical incorporation of the coupling of the \ac{scl} regions and regions outside of the \ac{scl} and add remarks on an efficient solution strategy for this system. Moreover, we present results computed with the proposed coupling approach to compare our solution with one-dimensional models, to find a quantitative measure for the size of the \ac{scl} region, to validate conservation principles, to quantify the quality of the proposed approach by defining approximation errors, and to show the applicability to large systems that represent realistic microstructures.\\
To our knowledge, this work is the first to show results for spatially resolved \acp{scl} within realistic microstructures.
\section*{Continuum model for solid electrolytes including \acp{scl}}
\label{sec:continuous_model}
In this section, we present a continuum model for \acp{scl} by splitting the geometry of the solid electrolyte into an \ac{scl} region close to the electrodes and the remaining domain. We summarize a thermodynamically consistent model for solid electrolytes, which is already established in the literature. Subsequently, we define different assumptions for the two domains and apply them to the thermodynamically consistent model. Afterwards, we elaborate on the coupling between both domains and discuss the approximation errors that we introduce by our proposed approach.
\subsection*{Geometric definitions and nomenclature}
\begin{figure}[ht]
    \centering
    \def\svgwidth{8.2cm}
    \fontsize{5}{10}\selectfont
    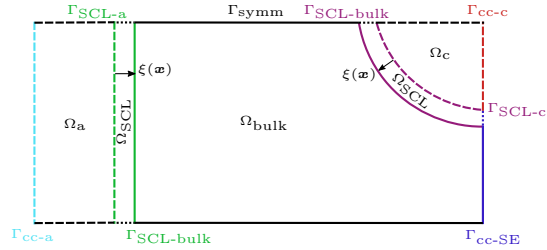
    \caption{Schematic sketch of the computational domain. The domain is split into subdomains~$\Omega_i$ and surfaces~$\Gamma_{i-j}$ denoting the interface between the domains~$\Omega_i$ and~$\Omega_j$.}
    \label{fig:geometry_domains_boundaries}
\end{figure}

Before presenting the equations that define the model for the solid electrolyte, we need to define the geometric setup as shown in \cref{fig:geometry_domains_boundaries} that schematically sketches both electrodes and the solid electrolyte.
The focus of this work is on the domain of the solid electrolyte~$\Omega_\text{SE}$ where the development of \acp{scl} is expected, as we do not consider charge separation in the electrodes in this work. The domain of the solid electrolyte is split into a part where we expect \acp{scl} to develop~$\Omega_\text{SCL}$ and into the bulk domain~$\Omega_\text{bulk}$: $\Omega_\text{SE} = \Omega_\text{SCL} \cup \Omega_\text{bulk}$. The electrodes~$\Omega_\text{ed}$ are subdivided into the anode~$\Omega_\text{a}$ and the cathode~$\Omega_\text{c}$: $\Omega_\text{ed} = \Omega_\text{a} \cup \Omega_\text{c}$. Their boundaries are drawn by dashed lines to indicate, that we do not solve any equations inside these domains within this work. Instead, we focus on the solid electrolyte as we are only interested in the \acp{scl} that form at the interface between the electrolyte and the electrodes. Thus, we do not resolve \acp{scl} at grain boundaries inside the solid electrolyte. However, their incorporation would be methodologically identical.\\
We define surfaces as intersections of domains or outer boundaries. At first, we define the intersection between the bulk domain of the solid electrolyte~$\Omega_\text{bulk}$ and the \ac{scl} domains~$\Omega_\text{SCL}$ as~$\Gamma_\text{SCL-bulk}$. The intersection between the electrodes~$\Omega_\text{ed}$ and the \ac{scl} domain~$\Omega_\text{SCL}$ is defined as~$\Gamma_\text{SCL-ed}$. All boundaries in the lateral direction of the battery cell are model boundaries, where symmetry assumptions are made and the according boundary conditions are applied. They are denoted with~$\Gamma_\text{symm}$. For completeness, we define outer boundaries to the current collectors~$\Gamma_\text{cc-a}$, $\Gamma_\text{cc-c}$, and~$\Gamma_\text{cc-SE}$ where~$\Gamma_\text{cc-SE}$ includes the boundaries of both the \ac{scl} domain~$\Omega_\text{SCL}$ and the bulk domain~$\Omega_\text{bulk}$ of the solid electrolyte. \\
Finally, a natural coordinate~$\xi(\vec{x})$ is introduced perpendicular to~$\Gamma_\text{SCL-bulk}$ which is restricted to ~$\Omega_\text{SCL}$. Its direction is defined from the electrodes to the electrolyte, with the origin at the electrode. Consequently, it is a function of the location~$\vec{x}$.
\subsection*{Thermodynamically consistent model for solid electrolytes including \acp{scl}}
\label{sec:scl_model}
We use the approach developed in~\cite{Braun2015}, and later extended in~\cite{BeckerSteinberger2021}, to model the electrochemical transport phenomena in solid electrolytes. This approach is thermodynamically consistent, i.e. a positive production rate of entropy is guaranteed. The key aspects of the approach as well as the used symbols are summarized in the Appendix \ref{sec:derivation_scl_model} and the governing equations are
\begin{alignat}{2}
    \parder{c_+}{t} + \nabla \cdot \vec{N}_{+} &= 0 &&\text{in} \ \Omega_\text{SE}, \label{eq:1D_scl_c+} \\
    \parder{q}{t} + \nabla \cdot \left (z_+ F \vec{N}_{+} -\epsilon_0 \chi \parder{\nabla \Phi}{t}\right ) &= 0 &&\text{in} \ \Omega_\text{SE} , \label{eq:1D_scl_q}\\
    -\nabla \cdot (\epsilon \nabla \Phi) &= q_{\text F}  &&\text{in} \ \Omega_\text{SE} , \label{eq:1D_scl_phi}\\
    \vec{N}_{+}  &= - D_+ \nabla c_+ - \frac{\sigma}{z_+ F} \nabla \Phi  \quad &&\text{in} \ \Omega_\text{SE} . \label{eq:1D_scl_N}
\end{alignat}
A physically meaningful (i.e.~no violation of conservation properties) and mathematically consistent set of boundary conditions needs to be defined to obtain a well-posed system. The shape of the \acp{scl} is determined by the boundary conditions applied to the interface between the solid electrolyte and the electrodes~$\Gamma_\text{SCL-ed}$. In the remaining sections of this work, we assume blocking electrodes which lead to Dirichlet boundary conditions for the electric potential and homogeneous Neumann boundary conditions for the concentration of cations on the interface
\begin{alignat}{2}
    \Phi &= \hat{\Phi} \quad  &&\text{on} \ \Gamma_\text{SCL-ed}, \\
    \vec{N}_+ \cdot \vec{n} &= 0  &&\text{on} \ \Gamma_\text{SCL-ed}.
\end{alignat}
For completeness, we define homogeneous Neumann boundary conditions as well for all quantities on~$\Gamma_\text{symm}$ and note that all boundary conditions have to satisfy global charge neutrality, namely $\int_{\Omega_\text{SE}} q(\vec{x}, t) \,\dd \Omega =0 \ \forall \ t$.\\
For the transient equations, we define initial conditions that represent an unpolarized solid electrolyte. This results in bulk concentration for anions~$c_-$ and cations~$c_+$, and thus implicitly zero total charge~$q$
\begin{align}
    c_+(\vec{x}, t=0) = c_-(\vec{x}, t=0) &= c_\text{bulk}, \\
    q(\vec{x}, t=0) &= 0.
\end{align}
\subsection*{Assumption of constant concentrations outside of the \ac{scl} region}
For the derivation of a model of the solid electrolyte outside of the \ac{scl} region, we start with the fundamental assumption
\begin{equation}
    \label{eq:bulk_assumption}
    c_+ = c_- = c_\text{bulk} = \text{const.} \quad \text{in} \ \Omega_\text{bulk}.
\end{equation}
This is a reasonable assumption for transference numbers close to unity if the local electroneutrality condition is satisfied~\cite{Bachman2015, Klerk2018}. As a direct consequence this assumption implies~$\frac{\partial c_+}{\partial t} = 0$ and $\nabla c_+ = \vec{0}$. Within this work, we keep the dielectric permeability independent of any excitation frequencies, i.e.~$\epsilon = \text{const}$. The assumption in~\cref{eq:bulk_assumption} ensures that no free charge (see~\cref{eq:free_charge}) accumulates, as positive and negative charges sum up to zero~$q_F = 0$.\\
In the following, we will show that those assumptions simplify the system of equations outlined before to a Laplace equation for the electric potential~$\Delta \Phi = 0$. First, we simplify the constitutive equation for the flux of cations in \cref{eq:1D_scl_N}. The diffusive term related to the gradient of the concentration vanishes
\begin{equation}
    \vec{N}_+ = - \frac{\sigma}{z_+ F} \nabla \Phi.
\end{equation}
Moreover, the transport properties become constants with respect to the cation concentration, namely the ionic conductivity and the diffusion coefficient
\begin{align}
    \sigma &= (z_+ F)^2 \mathcal{L_{++}} (1-(c_{\text{max}} - c_\text{bulk}) c_\text{bulk} \Delta \nu), \\
    D_+ &= \mathcal{L_{++}} R T \frac{c_{\text{max}}}{(c_{\text{max}} -c_\text{bulk}) c_\text{bulk}}.
\end{align}
By using the absence of free charge~$q_\text{F}$, \cref{eq:1D_scl_phi} simplifies to $\nabla \cdot (\epsilon \nabla \Phi) = 0$, and by using constant dielectric permeability $\epsilon$ it simplifies further to the Laplace equation
\begin{equation}
    \Delta \Phi = 0 \quad \text{in} \ \Omega_\text{bulk}.
\end{equation}
Furthermore, \cref{eq:1D_scl_c+} reduces to $\nabla \cdot \left( - \frac{\sigma}{z_+ F} \nabla \Phi \right) = 0$ by using that the temporal derivative of the cation concentration is zero and by substituting the expression for the flux of cations $\vec{N}_+$. All prefactors are constant in this expression, such that this equation reduces to $\Delta \Phi = 0$ as well.\\
Finally, we substitute all findings into \cref{eq:1D_scl_q} and apply the divergence operator on both terms inside the brackets
\begin{equation}
    \parder{q}{t} + \left (- \sigma \Delta \Phi - \epsilon_0 \chi \parder{\Delta \Phi}{t} \right) = 0.
\end{equation}
As shown before, the Laplacian of~$\Phi$ will evaluate to zero, such that only the temporal derivative of the total charge density $\parder{q}{t}$ remains. The total charge density~$q$ is composed by summing up the free charge density~$q_F$ and the bound charge density~$q_\text{B}$. For the first, we already know that it is zero, while the latter is defined as~$q_B = \epsilon_0 \chi \Delta \Phi$. Again, we make use of the Laplacian of~$\Phi$ to be zero leading to~$q_\text{B} = 0$. Finally, we can conclude that~$q=0$. Obviously, this equation ($0=0$) is implicitly fulfilled. This means, that from the entire set of equations, we only need to solve for the Laplacian of~$\Phi$ to be zero in~$\Omega_\text{bulk}$.
\subsection*{Assumption of one-dimensionality inside the \ac{scl} region}
\label{sec:set_equations}
The key aspect of this work is to propose a model that can spatially resolve the effect of \acp{scl} in realistic microstructures by reducing the required computational effort. We do this by assuming, that all spatial derivatives inside $\Omega_\text{SCL}$ that are tangential to the interface $\Gamma_\text{SCL-bulk}$ vanish
\begin{equation}
    (\nabla \Psi - (\nabla \Psi \cdot \vec{n}) \ \vec{n}) \cdot \ \vec{n} = 0 \quad \text{in} \ \Omega_\text{SCL},
\end{equation}
for any scalar quantity $\Psi$ and the vector $\vec{n}$ being normal to the interface $\Gamma_\text{SCL-bulk}$ with length one. This assumption is motivated by the observable main characteristics of \acp{scl} that are predominantly one-dimensional. The one-dimensionality is caused by the electric field $\vec{E}= - \nabla \Phi$ which has to be perpendicular to equipotential, i.e.~ideally conducting surfaces $\nabla \Phi \cdot \vec{n} = \lVert \nabla \Phi \rVert$ as a limit assumption for the high conductivity of many electrode materials. Furthermore, we distinguish between the steady state and the transient state for blocking electrodes. In the steady state, the flux of cations vanishes, and \cref{eq:1D_scl_N} simplifies to $\nabla c_+ = -\frac{\sigma}{D_+z_+ F} \nabla \Phi$. Consequently, the gradient of the cation concentration in the steady state is normal to the surface as well. Thus, the electric potential and the concentration are constant on this surface. Now, we conclude that the electric potential and the concentration at an infinitesimal distance from the surface are equal as well, due to the constant values at the surface and normal gradients with uniform magnitude. This surface with an infinitesimal distance forms another surface with uniform electric potential and concentration. Repeating this thought experiment reveals that the gradient of the electric potential and the concentration is normal to the surface throughout the entire \ac{scl} domain, i.e. that all gradients remain parallel to the normal of the surface.\\
In the transient state, the tangential component of the flux of cations can have non-zero values but remain comparably small as we will show in this work.\\
Consequently, the partial differential equations as outlined before are reduced to one-dimensional equations. Inside the remaining part of the geometrically complex solid electrolyte $\Omega_\text{bulk}$, no further constraint to the gradient is given, such that the equations are resolved in all three dimensions of space. Considering this, we arrive at a set of equations inside both the \ac{scl} domain $\Omega_\text{SCL}$ and the bulk domain $\Omega_\text{bulk}$
\begin{alignat}{2}
    \Delta \Phi_\text{bulk} &= 0 &&\text{in} \ \Omega_\text{bulk}, \label{eq:set_equations_first}\\
    \parder{c_\text{SCL}}{t} + \parder{N_\text{SCL}}{\xi} &= 0 &&\text{in} \ \Omega_\text{SCL}, \label{eq:set_equations_second}\\
    \parder{q_\text{SCL}}{t} + \parder{\left (z F N_\text{SCL} -\epsilon_0 \chi \parder{\left(\parder{\Phi_\text{SCL}}{\xi}\right)}{t}\right)}{\xi} &= 0 &&\text{in} \ \Omega_\text{SCL}, \label{eq:set_equations_third}\\
    -\parder{(\epsilon \parder{\Phi_\text{SCL}}{\xi})}{\xi} &= q_{\text F} \quad &&\text{in} \ \Omega_\text{SCL},\label{eq:set_equations_fourth}\\
    N_\text{SCL} &= - D \parder{c_\text{SCL}}{\xi} - \frac{\sigma }{z F} \parder{\Phi_\text{SCL}}{\xi}, \label{eq:set_equations_last}
\end{alignat}
where we abbreviated the cation concentration~$c_+$ with~$c$, the flux of cations~$N_+$ with~$N$, the charge number~$z_+$ with~$z$, and the diffusion coefficient~$D_+$ with~$D$, as from now on the cation concentration mathematically is the only unknown concentration. Additionally, we distinguish between quantities in the bulk domain and in the \ac{scl} domain by assigning the respective subscript. In the following, we do not solve for the total charge~$q_\text{SCL}$ in \cref{eq:set_equations_third}, as it is not an independent variable in the case of blocking electrodes and can simply be post-processed from the electric potential and the cation concentration.
\subsection*{Coupling regions inside and outside of the \ac{scl}}
\label{sec:coupling_conditions}
At the transition from the bulk domain to the \ac{scl} domain $\Gamma_\text{SCL-bulk}$ we require continuity between all primary variables, i.e.~the concentration, and the electric potential. Additionally, conservation properties need to be ensured. The first requirement is fulfilled by requesting
\begin{alignat}{2}
    \Phi_\text{SCL} &= \Phi_\text{bulk} \quad &&\text{on} \ \Gamma_\text{SCL-bulk}, \\
    c_\text{SCL} &= c_\text{bulk} &&\text{on} \ \Gamma_\text{SCL-bulk}.
\end{alignat}
The second requirement can be incorporated by enforcing consistent coupling fluxes between two domains. While the flux inside the three-dimensional bulk domain is a vector, the flux inside the one-dimensional \ac{scl} domain is treated as a scalar. Thus, the flux inside the bulk domain needs to be projected in the direction normal to the interface implying the one-dimensional \ac{scl} domain to be perpendicular to the coupling surface $\Gamma_\text{SCL-bulk}$
\begin{alignat}{2}
    \vec{N}_\text{bulk} \cdot \vec{n} &= N_\text{SCL} \quad &&\text{on} \ \Gamma_\text{SCL-bulk}, \\
    \vec{i}_\text{bulk} \cdot \vec{n} &= i_\text{SCL} &&\text{on} \ \Gamma_\text{SCL-bulk}.
\end{alignat}
\subsection*{Geometric approximation of realistic microstructures}
We define the bulk domain of the solid electrolyte as~$\Omega_\text{bulk} = \Omega_\text{SE}$ and the \ac{scl} domain as~$\Omega_\text{SCL} = \Gamma_\text{SCL-bulk} \times l_\text{SCL}$. Here,~$l_\text{SCL}$ is an estimation of the thickness of the \ac{scl} and~$l_\text{SCL} << l_\text{SE}$, with~$l_\text{SE}$ being a typical length scale of the solid electrolyte. Note, that this approximation slightly enlarges the original geometry.
\subsection*{Approximation errors introduced by the coupling approach}
\label{sec:errors_by_coupling}
By coupling the one-dimensional and the three-dimensional domains, we introduce three types of approximation errors to the system. They serve as a measure to quantify the quality of the proposed approach:
\begin{enumerate}
    \item Model error. The model for the bulk domain is derived from the thermodynamically consistent model for \acp{scl} based on the assumption~$c_+ = c_- = c_\text{bulk}$. Thus, the error introduced by the assumption scales with~$\epsilon_\text{err} = c_+ - c_\text{bulk}$. It is negligible if the \ac{scl} domain is chosen large enough as the concentration converges towards the bulk concentration for great distances from~$\Gamma_\text{SCL-ed}$.
    \item Geometric error. By adding a thin layer representing the \ac{scl} domain we modify the geometry and thus enlarge the geometric dimensions of the solid electrolyte. Effectively, this results in a slightly larger resistance of the solid electrolyte. However, we select the thickness of the additional layer~$l_\text{SCL} << l_\text{SE}$ which means that the additional resistance, which scales with the length of the solid electrolyte, is negligible. In case the aforementioned condition is not valid anymore, it is possible to reduce the size of the bulk domain to compensate for the additional thin layer representing the \ac{scl}.
    \item Compatibility error. The one-dimensional model inside the \ac{scl} domain can only capture gradients in the direction normal to~$\Gamma_\text{SCL-bulk}$. Gradients parallel to~$\Gamma_\text{SCL-bulk}$ on equipotential surfaces occur in the transient state but cannot cause a flux in the one-dimensional model. Again, this error is comparably small as long as~$l_\text{SCL} << l_\text{SE}$. Exemplarily, this can be shown by a Taylor expansion of the electric potential for a two-dimensional geometry in polar coordinates ($r$, $\theta$) to capture the curvature of the equipotential surface:
    \begin{equation}
        \Phi(r,\theta) = \Phi(\vec{0}) +
        \left.
        \begin{bmatrix}
            \parder{\Phi}{r} & \frac{1}{r} \parder{\Phi}{\Theta}
        \end{bmatrix}
        \right|_{\vec{0}}
        \begin{bmatrix}
            r \\
            \theta
        \end{bmatrix}
        + \frac{1}{2}
        \begin{bmatrix}
            r & \theta
        \end{bmatrix}
        \left.
        \begin{bmatrix}
            \frac{\partial^2 \Phi}{\partial r^2} & \frac{\partial^2 \Phi}{\partial r \partial \theta} - \frac{1}{r}\parder{\Phi}{\theta} \\
            \frac{\partial^2 \Phi}{\partial r \partial \theta} - \frac{1}{r}\parder{\Phi}{\theta} & \frac{\partial^2 \Phi}{\partial \theta^2} + \frac{1}{r}\parder{\Phi}{r}
        \end{bmatrix}
        \right|_{\vec{0}}
        \begin{bmatrix}
            r \\
            \theta
        \end{bmatrix}
        + \mathcal{O}^3(r, \theta)
    \end{equation}
    Due to the equipotential surface, the derivatives $\parder{\Phi}{\theta}$ and $\frac{\partial^2 \Phi}{\partial \theta^2}$ vanish. The compatibility error scales with derivatives the reduced dimensional model cannot capture, i.e.~derivatives w.r.t.~$\theta$, and thus w.r.t.~$\left. \frac{\partial^2 \Phi}{\partial r \partial \theta} \right|_{\vec{0}} r \theta$. Therefore, the distance to the equipotential surface must be minimal to reduce the compatibility error.
\end{enumerate}
We conclude, that the domain of the \ac{scl} should be as large as possible to reduce the first approximation error, while it should be as small as possible, to reduce the other two approximation errors. We will present concepts on how to choose the size of the \ac{scl} domain.
\section*{Numerical treatment of the \ac{scl} model}
\label{sec:numerics}
In this section we want to present the discretization schemes in space and time we used to discretize the continuous equations. Furthermore, we show the incorporation of the coupling conditions between the bulk domain and the \ac{scl} domain into the discretized system of equations. Afterwards, we discuss the required constraint enforcement and the applied solution techniques.
\subsection*{Discretization in time}
The equation for the bulk domain is stationary, while the set of equations for the \ac{scl} domain contains time derivatives, namely the temporal derivative of the concentration. Thus, only the mass conservation equation in the \ac{scl} domains is discretized in time. It is not the aim of this work to rewrite in all detail the steps for discretizing the time-continuous equations. For brevity, the main steps of the One-Step-Theta method which is used in this work are recaptured. It is used to discretize first-order differential equations of the type
\begin{equation}
    \parder{c_\text{SCL}}{t} = \text{fn}(c_\text{SCL}, \vec{x}),
\end{equation}
in time for $t \in[t_0, t_\text{end}]$. The underlying discretization scheme can be expressed as
\begin{equation}
    \parder{c_\text{SCL}}{t} \approx \frac{c_\text{SCL}^{n+1} - c_\text{SCL}^{n}}{\Delta t} = \theta \ \text{fn}(c_\text{SCL}^{n+1}, \vec{x}) + (1 - \theta) \ \text{fn}(c_\text{SCL}^n, \vec{x}),
\end{equation}
with~$c_\text{SCL}^n$ and~$c_\text{SCL}^{n+1}$ being the values of~$c_\text{SCL}$ at time steps~$t_n$ and~$t_{n+1}$, with possibly non-uniform values of the time step size~$\Delta t$. Choosing~$\theta = 0.5$ represents the well-established implicit Crank-Nicolson scheme of second-order accuracy.
\subsection*{Discretization in space}
We use the finite element method to discretize the set of partial differential equations in space for both the bulk domain $\Omega_\text{bulk}$ and the \ac{scl} domain $\Omega_\text{SCL}$. While the bulk domain is discretized in all three dimensions of space, the domain for the \ac{scl} is only discretized in one dimension. Consequently, the number of unknowns of the spatial discretization reduces significantly, as a fine discretization in only one direction is required. Again, we do not aim to walk the reader through all steps of the discretization scheme of the finite element method but want to focus on the main aspects. First, we derive the weak form of \cref{eq:set_equations_first,eq:set_equations_second,eq:set_equations_third,eq:set_equations_fourth,eq:set_equations_last} by multiplication with an arbitrary test function~$w$, integration over the respective domain, and transforming derivatives of second order in space to the test function~$w$ by applying the chain rule of divergence and Gau\ss \ divergence theorem. Afterwards, we discretize the geometry ($\vec{x}$), the test functions ($w$), and the solution variables ($\Phi_\text{bulk}$, $c_\text{SCL}$, $\Phi_\text{SCL}$) with the same shape functions. This means that $\vec{\Psi} = \mat{N} \vec{\hat{\Psi}}$, with~$\vec{\Psi} \in \{ \vec{x}, w, \Phi_\text{bulk}, c_\text{SCL}, \Phi_\text{SCL} \}$ represents the vector of all variables, $\mat{N}$ the matrix of corresponding size containing the shape functions, and $\vec{\hat{\Psi}}$ the vector of the discretized variables. Throughout this work, we use linear shape functions for the matrix $\mat{N}$. Finally, we arrive at a set of nonlinear, algebraic equations
\begin{align}
    \vec{R}_\Phi^\text{bulk} (\vec{\Phi}_\text{bulk}^{n+1}) = \vec{0}, \\
    \vec{R}_c^\text{SCL} (\vec{c}_\text{SCL}^{n+1}, \vec{\Phi}_\text{SCL}^{n+1}) = \vec{0}, \\
    \vec{R}_\Phi^\text{SCL} (\vec{c}_\text{SCL}^{n+1}, \vec{\Phi}_\text{SCL}^{n+1}) = \vec{0},
\end{align}
where $\vec{\Phi}_\text{bulk}^{n+1}$, $\vec{c}_\text{SCL}^{n+1}$, and $\vec{\Phi}_\text{SCL}^{n+1}$ denote the vector-organized nodal values of the primary variables. 
\subsection*{Solution of the algebraic nonlinear system of equations}
The set of nonlinear equations is solved by the Newton-Raphson scheme. Both, the primary variables and the residuals are combined into global vectors $\vec{\omega}^{n+1} = [\vec{\Phi}_\text{bulk}^{n+1}, \vec{c}_\text{SCL}^{n+1}, \vec{\Phi}_\text{SCL}^{n+1}]^\text{T}$ and $\vec{R} = [\vec{R}_\Phi^\text{bulk}, \vec{R}_c^\text{SCL}, \vec{R}_\Phi^\text{SCL}]^\text{T}$. This allows defining the Newton-Raphson scheme as
\begin{equation}
    \vec{\omega}_{i+1}^{n+1} = \left( \left. \parder{\vec{R}}{\vec{\omega}^{n+1}} \right|_i \right)^{-1} \vec{R}_i + \vec{\omega}_i^{n+1},
\end{equation}
where the inverse of the matrix $\left. \parder{\vec{R}}{\vec{\omega}^{n+1}} \right|_i$ is of course not computed. Instead, the system $\left. \parder{\vec{R}}{\vec{\omega}^{n+1}} \right|_i \Delta \vec{w}_{i+1}^{n+1} = \vec{R}_i$ is solved using a linear solver, with $\Delta \vec{w}_{i+1}^{n+1} = \vec{w}_{i+1}^{n+1} - \vec{w}_{i}^{n+1}$. The matrix $\left. \parder{\vec{R}}{\vec{\omega}^{n+1}} \right|_i$ can be written as
\begin{equation}
    \renewcommand\arraystretch{2}
    \left.
    \begin{bmatrix}
        \parder{\vec{R}_\Phi^\text{bulk}}{\vec{\Phi}_\text{bulk}} && \mat{0} && \mat{0} \\
        \mat{0} && \parder{\vec{R}_c^\text{SCL}}{\vec{c}_\text{SCL}} && \parder{\vec{R}_c^\text{SCL}}{\vec{\Phi}_\text{SCL}} \\
        \mat{0} && \parder{\vec{R}_\Phi^\text{SCL}}{\vec{c}_\text{SCL}} && \parder{\vec{R}_\Phi^\text{SCL}}{\vec{\Phi}_\text{SCL}} \\
    \end{bmatrix}
    \right|_i
    \renewcommand\arraystretch{1}
    =
    \left.
    \begin{bmatrix}
        \mat{K}^\text{bulk}_{\Phi,\Phi} && \mat{0} && \mat{0} \\
        \mat{0} && \mat{K}^\text{SCL}_{c,c} && \mat{K}^\text{SCL}_{c,\Phi} \\
        \mat{0} && \mat{K}^\text{SCL}_{\Phi,c} && \mat{K}^\text{SCL}_{\Phi,\Phi} \\
    \end{bmatrix}
    \right|_i
    = \mat{K}_i.
\end{equation}
For brevity, we summarize the expressions for the single blocks of the matrix by introducing submatrices~$\mat{K}^\Omega_{\Psi_1,\Psi_2}$.
\subsection*{Incorporation of the coupling by constraint enforcement}
\label{sec:coupling_scheme}
\begin{figure}[ht]
    \centering
    \def\svgwidth{8.2cm}
    \fontsize{5}{10}\selectfont
    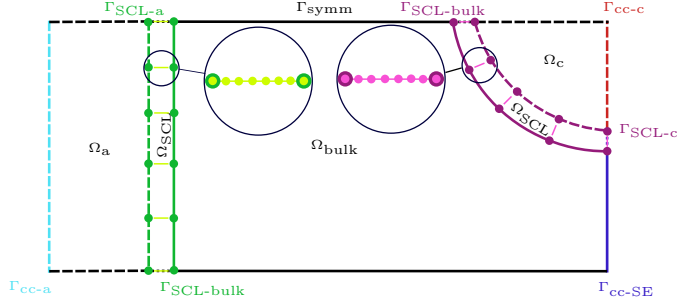
    \caption{Schematic sketch of the discrete coupling between the two domains with different dimensions of space.}
    \label{fig:geometry_domains_boundaries_discrete}
\end{figure}

\begin{figure}[ht]
    \centering
    \def\svgwidth{4.0cm}
    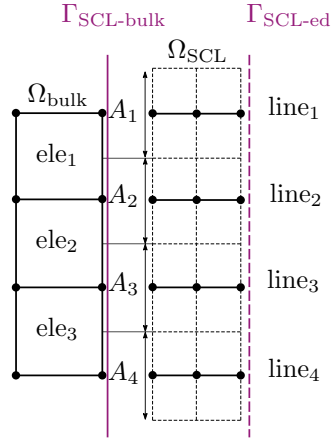  
    \caption{Projected one-dimensional discretization on the two-dimensional interface. Scaling of the equations in the \ac{scl} domain with the projected area $A_i$ guarantees the conservation properties across the interface $\Gamma_\text{SCL-bulk}$.}
    \label{fig:discrete_coupling}
\end{figure}

In \cref{fig:geometry_domains_boundaries_discrete} we show the discrete coupling scheme between the bulk domain and the \ac{scl} domain.
The bulk domain $\Omega_\text{bulk}$ is discretized with standard finite elements for three-dimensional spaces, meaning hexahedrals or tetrahedrals. By their choice, the surface discretization of the interface $\Gamma_\text{SCL-bulk}$ is determined, namely by quadrilaterals and triangulars. We connect the nodes on $\Gamma_\text{SCL-bulk}$ and $\Gamma_\text{SCL-ed}$ by introducing a one-dimensional discretization consisting of line elements to discretize~$\Omega_\text{SCL}$. The mesh of the one-dimensional discretization can be much finer compared to the three-dimensional mesh in $\Omega_\text{bulk}$.\\
The coupling conditions derived before will now be imposed on the linear system of equations to couple the three-dimensional discretization in~$\Omega_\text{bulk}$ with the one-dimensional discretization in $\Omega_\text{SCL}$. At first, we consider the requirement of conservation across the coupling interface. Therefore, we assign an area $A_i$ to each one-dimensional \ac{scl} discretization to extend the discretization to all three dimensions of space (see \cref{fig:discrete_coupling}) which is consistent with the chosen linear shape functions and could be extended to higher-order shape functions.
For the linear shape functions, this corresponds to a piecewise constant behavior of the \ac{scl} domain in the tangential direction of the interface~$\Gamma_\text{SCL-bulk}$. Subsequently, we scale both the residual~$\vec{R}^\text{SCL}$ and the linearization matrix~$\mat{K}^\text{SCL}$ of the \ac{scl} domain with the projected areas~$A_i$. We organize the projected areas~$A_i$ in a vector and evaluate them by integrating the shape functions at the interface over the constant value one: $\vec{A} = \Aop_\text{ele} \int_{\Gamma_\text{coup,ele}} \mat{N} \text{d} \Gamma$, with the assembly operator~$\Aop_\text{ele}$. This results in~$\vec{R}^\text{SCL,coup}_\Psi = \vec{R}^\text{SCL}_\Psi . \vec{A}$ and~$\mat{K}_{\Psi_1,\Psi_2}^\text{SCL, coup} = \mat{K}_{\Psi_1,\Psi_2}^\text{SCL} . \vec{A}$, with '$.$' denoting the operator for row-wise multiplication.\\
Now, we can enforce the requirement of continuity between the bulk domain and the SCL domain. This is achieved by (a) splitting the primary variables into coupled variables $\Psi_\text{coup}$ and interior variables $\Psi_\text{i}$, (b) further subdividing the coupled variables in accordance with conventions into "slave" on the \ac{scl} side and "master" on the bulk side, labeled with "s" and "m" respectively, (c) introducing Lagrangian multipliers to enforce the constraint of continuity $\vec{\Psi}_\text{bulk} = \vec{\Psi}_\text{SCL}$ at $\Gamma_\text{SCL-bulk}$, and (d) applying a condensation scheme to remove the Lagrangian multipliers as well as the slave-side interface variables from the system of equations defining the final linear system of equations
\begin{equation}
    \renewcommand\arraystretch{1.2}
    \scriptstyle
    \left.
    \begin{bmatrix}
        \mat{K}^\text{bulk}_{\Phi_\text{i},\Phi_\text{i}} && \mat{K}^\text{bulk}_{\Phi_\text{i},\Phi_\text{m}} &&  \mat{0} && \mat{0} \\
        \mat{K}^\text{bulk}_{\Phi_\text{m},\Phi_\text{i}} && \mat{K}^\text{bulk}_{\Phi_\text{m},\Phi_\text{m}} + \mat{K}^\text{SCL}_{\Phi_\text{s},\Phi_\text{s}} . \vec{A} && \mat{K}^\text{SCL}_{\Phi_\text{s},c} . \vec{A} && \mat{K}^\text{SCL}_{\Phi_\text{s},\Phi_\text{i}} . \vec{A} \\
        \mat{0} && \mat{K}^\text{SCL}_{c,\Phi_\text{s}} . \vec{A} && \mat{K}^\text{SCL}_{c,c} . \vec{A}  && \mat{K}^\text{SCL}_{c,\Phi_\text{i}} . \vec{A} \\
        \mat{0} && \mat{K}^\text{SCL}_{\Phi_\text{i},\Phi_\text{s}} . \vec{A} && \mat{K}^\text{SCL}_{\Phi_\text{i},c} . \vec{A} && \mat{K}^\text{SCL}_{\Phi_\text{i},\Phi_\text{i}} . \vec{A} \\
    \end{bmatrix}
    \right|_i
    \left.
    \begin{bmatrix}
        \Delta \vec{\Phi}_\text{i}^\text{bulk} \\
        \Delta \vec{\Phi}_\text{m}^\text{bulk} \\
        \Delta \vec{c}^\text{SCL} \\
        \Delta \vec{\Phi}_\text{i}^\text{SCL} \\
    \end{bmatrix}
    \right|_{i+1}
    =
    \left.
    \begin{bmatrix}
        \vec{R}_{\Phi_\text{i}}^\text{bulk} \\
        \vec{R}_{\Phi_\text{m}}^\text{bulk} + \vec{R}_{\Phi_s}^\text{SCL} . \vec{A} \\
        \vec{R}_c^\text{SCL} . \vec{A} \\
        \vec{R}_{\Phi_\text{i}}^\text{SCL} . \vec{A}
    \end{bmatrix}
    \right|_i
    .
    \renewcommand\arraystretch{1}
\end{equation}
\subsection*{Solution of the linearized system of equations}
Within this work, we choose a monolithic coupling scheme to solve the outlined linear system of equations where equations from the bulk domain and the \ac{scl} domain are coupled. As shown elsewhere (e.g.~\cite{Verdugo2016} for n-field problems or \cite{Fang2018} for electrochemical problems), the monolithic solution approach is seen as superior considering robustness and often also with respect to efficiency compared to other schemes like partitioned coupling or sub-cycling for various types of applications. The most prominent drawback of the monolithic coupling approach is, that the underlying matrix is comparably ill-conditioned. This is caused by additional entries in the matrix that are far away from the main diagonal, and entries with different orders of magnitude originating from the different domains, dimensions, and discretization coarseness that are coupled. Thus, standard iterative solvers that are required to solve realistic microstructures with a large number of unknowns, are not applicable anymore and tailored preconditioners are required. We choose a combined Block-Gau\ss-Seidel and Algebraic-Multigrid preconditioner as outlined in \cite{Fang2019}. The core idea is to split the full linear system of equations into subblocks, that are physically meaningful, e.g.~geometric domains or types of primary variables, and apply a Block-Gau\ss-Seidel scheme on these blocks. This already improves the condition of the subblocks compared to the full system of equations. Additionally, we perform a prescaling of the rows and the columns of the subblocks to further improve the condition of the subblocks. Finally, we apply an Algebraic-Multigrid preconditioner to the subblocks on the main diagonal within the Block-Gau\ss-Seidel iteration.
\section*{Results}
\label{sec:results}
The results presented in this section are computed with BACI~\cite{BACI}, our in-house multi-physics research code. We begin with approximating an optimal length and discretization size for the \ac{scl} domain, validating the proposed model, and conclude with showing the applicability of the model to realistic microstructures.
\subsection*{Materials}
The idea of this paper is not to investigate the behavior of the \ac{scl} for different materials and conditions. Instead, we want to analyze the proposed model in more detail. Hence, we restrict ourselves to one set of material parameters (for lithium lanthanum titanate - LLTO) throughout this work if not explicitly stated to be different. All relevant material parameters, initial conditions, and physical constants are chosen as in \cite{BeckerSteinberger2021} and are listed in \cref{table:material_params_init_conds}.
\begin{table}[ht]
    \renewcommand{\arraystretch}{1.1}
    \centering
    \caption{Material parameters of LLTO, initial and boundary conditions, and natural constants for all simulations.}
    \begin{tabular}{c | c | c | c}
      \textbf{quantity}                             & \textbf{symbol}   & \textbf{value}     & \textbf{source}                \\
      \hline
        ionic conductivity                 & $\sigma$                                     & $0.02 \ \frac{\text{S}}{\text{m}}$                & \cite{Kwon2017} \\
        maximal concentration              & $c_\text{max}$                               & $14214 \ \frac{\text{mol}}{\text{m}^3} $         & calculated \\
        bulk concentration                 & $c_\text{bulk}$                              & $9476 \ \frac{\text{mol}}{\text{m}^3}$           & calculated \\
        lower bound of bulk concentration  & $c_\text{bulk,min} = 0.999 \ c_\text{bulk} $ & $9466.5 \ \frac{\text{mol}}{\text{m}^3}$         & defined \\
        upper bound of bulk concentration  & $c_\text{bulk,max} = 1.001 \ c_\text{bulk}$  & $9485.5 \ \frac{\text{mol}}{\text{m}^3}$         & defined \\
        tolerance of concentration         & $c_{\epsilon}$                               & $10^{-4} \frac{\text{mol}}{\text{m}^3}$           & defined \\
        susceptibility in SCLs             & $\chi$                                       & $10^5$                                            & \cite{Bucheli2014} \\
        difference in partial molar volume & $\Delta \nu = \nu_+ - \nu_v $                & $0$                                               & \cite{Braun2015} \\
        transference number of cations     & $t_+$                                        & $1$                                               & defined \\
        charge number                      & $z$                                          & $1$                                               & \cite{BeckerSteinberger2021} \\
        average molar mass                 & $M$                                          & $0.168822 \ \frac{\text{kg}}{\text{mol}}$         & calculated in \cite{BeckerSteinberger2021} \\
        average mass density               & $\rho$                                       & $4000 \ \frac{\text{kg}}{\text{m}^3}$            & \cite{Jena2005} \\
        \hline
        initial concentration              & $c_0$                                        & $9476 \frac{\text{mol}}{\text{m}^3}$             & equals $c_\text{bulk}$ \\
        difference in potential            & $\Delta \Phi$                                & $2 \ \text{V}$                                    & defined, as in \cite{BeckerSteinberger2021} \\
        temperature                        & $T$                                          & $298 \ \text{K}$                                  & defined, as in \cite{BeckerSteinberger2021} \\
        \hline
        dielectric permittivity of vacuum  & $\epsilon_0$                                 & $8.85 \cdot 10^{-12} \ \frac{\text{F}}{\text{m}}$ & defined, as in \cite{BeckerSteinberger2021} \\
        Faraday constant                   & $F$                                          & $9.65 \cdot 10^4 \ \frac{\text{C}}{\text{mol}}$   & defined, as in \cite{BeckerSteinberger2021} \\
        universal gas constant             & $R$                                          & $8.314 \ \frac{\text{J}}{\text{mol K}}$           & defined, as in \cite{BeckerSteinberger2021}
    \end{tabular}
    \label{table:material_params_init_conds}
\end{table}
\subsection*{Two characteristic values to quantify an \ac{scl}}
\label{sec:characteristic_values}
We compute two characteristic quantities to quantify \acp{scl}: the spatial thickness~$d_\text{SCL}(\vec{x})$ and the integrated free charge~$Q_\text{SCL}$. By using fixed values for $c_\text{bulk,min}$ and $c_\text{bulk,max}$ we can define the thickness of the \ac{scl} by
\begin{equation}
    d_\text{SCL}(\vec{x}) = \xi(\vec{x}) \ \text{where} \left\{
    \begin{array}{ll}
        \text{argmin}(c(\xi(\vec{x})) > c_\text{bulk,min}) & \quad \text{if} \ c < c_\text{bulk} \\
        \text{argmin}(c(\xi(\vec{x})) < c_\text{bulk,max}) & \quad \textrm{else} \\
    \end{array}
    \right.
    .
\end{equation}
The integrated deviation from the neutrally charged state~$Q_\text{SCL}$ is computed as the integrated difference of the concentration~$c$ from the bulk concentration~$c_\text{bulk}$ scaled by the charge number~$z$ and Faraday's constant~$F$ to obtain a charge
\begin{equation}
    Q_\text{SCL} = z F \int_{\Omega_\text{SCL}} \left( c(\vec{x}) - c_\text{bulk} \right) \dd \Omega,
\end{equation}
where~$\Omega_\text{SCL}$ can be divided into the part at the anode~$\Omega_\text{SCL,a}$ and at the cathode~$\Omega_\text{SCL,c}$, and subsequently the integrated values~$Q_{\text{SCL,a}}$ and~$Q_{\text{SCL,c}}$, respectively. From the conservation of mass and charge we know, that a consistent formulation needs to fulfill~$Q_{\text{SCL,a}} + Q_{\text{SCL,c}} = Q_\text{SCL} = 0 \ \forall \ t$.
\subsection*{Suitable representation of the \ac{scl} domain}
We estimate the optimal length and discretization size for the \ac{scl} domain within the coupled model based on simulations with a one-dimensional model as outlined before and the parameters from \cref{table:material_params_init_conds}.
\subsubsection*{Optimal length of the \ac{scl} domain}
\begin{figure}[ht]
    \centering
    \begin{subfigure}{0.49\textwidth}
        \centering
        \vfill
        \input{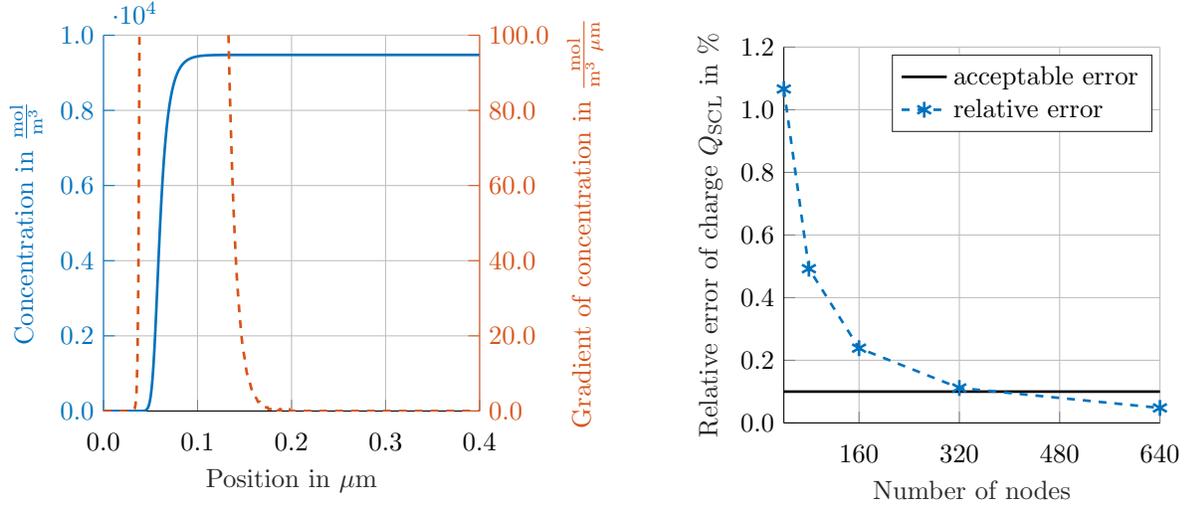}
        \caption{Concentration and gradient of the concentration show that $l_\text{SCL} = 0.4 \ \mu \text{m}$ is a good choice for the length of the \ac{scl} domain for the used material parameters and boundary conditions.}
        \label{fig:validate_extension}
        \vfill
    \end{subfigure}
    \hfill
    \begin{subfigure}{0.49\textwidth}
        \centering
        \vfill
        \definecolor{mycolor1}{rgb}{0.00000,0.44700,0.74100}%
\begin{tikzpicture}
\begin{axis}[%
width=5.0cm,
height=5.0cm,
scale only axis,
xmin=40,
xmax=640,
xtick={0,160,320,480,640},
x tick label style={
/pgf/number format/.cd,
fixed,
fixed zerofill,
precision=0,
/tikz/.cd,
yshift=-.5em},
xlabel style={font=\color{white!15!black}},
xlabel={Number of nodes},
ymin=0,
ymax=1.2,
ytick={0,0.2,0.4,0.6,0.8,1.0,1.2},
y tick label style={
/pgf/number format/.cd,
fixed,
fixed zerofill,
precision=1,
/tikz/.cd},
ylabel style={font=\color{white!15!black}},
ylabel={Relative error of charge $Q_\text{SCL}$ in $\%$},
axis background/.style={fill=white},
axis x line*=bottom,
axis y line*=left,
xmajorgrids,
ymajorgrids,
legend style={legend cell align=left, align=left, draw=white!15!black}
]
\addplot [color=black, line width=1.0pt]
  table[row sep=crcr]{%
40	0.1\\
640	0.1\\
};
\addlegendentry{acceptable error}
\addplot [color=mycolor1, line width=1.0pt,dashed, mark=asterisk, mark size=3pt, mark options={solid, color=mycolor1}]
  table[row sep=crcr]{%
40	1.06624783556804\\
80	0.492517427731681\\
160	0.238648895453571\\
320	0.111652253753759\\
640	0.0479044306608335\\
};
\addlegendentry{relative error}
\end{axis}
\end{tikzpicture}%
        \caption{Relative error of integrated charge $Q_\text{SCL}$ in the steady state for the used material parameters and boundary conditions.}
        \label{fig:charge_error}
        \vfill
    \end{subfigure}
    \caption{Determining the optimal length and discretization size for the \ac{scl} domain.}
\end{figure}

While the geometric representation of the bulk domain $\Omega_\text{bulk}$ is already defined by the problem statement, the extension $l_\text{SCL}$ of the \ac{scl} domain~$\Omega_\text{SCL}$ has to be determined for the proposed coupling approach. We select it based on the following two criteria to minimize the approximation errors defined before:
\begin{enumerate}
    \item The length of the \ac{scl} domain~$l_\text{SCL}$ must be large enough to enable the complete formation of \acp{scl} at interfaces of the electrodes with the solid electrolyte $\Gamma_{\text{SCL-bulk}}$. This can be expressed in the requirement of vanishing gradients of the concentration $\nabla c (\xi = l_\text{SCL}) = 0 \, \forall \ t$ in $\Omega_\text{SCL}$ in accordance with \cref{eq:bulk_assumption}.
    \item The length of the \ac{scl} domain~$l_\text{SCL}$ must be as small as possible to minimize the geometric error and the compatibility error.
\end{enumerate}
We combine two findings established in the literature to estimate a value of $l_\text{SCL}$:
\begin{enumerate}
    \item The ratio $\frac{l_\text{SCL}}{l_\text{SE}}$ is proportional to a non-dimensional length scale $\lambda$, which is similar to the Debye-length of double layers in liquid electrolytes~\cite{Braun2015} and defined as 
\begin{equation}
    \lambda = \sqrt{ \frac{k_\text{B} T \epsilon M}{e_0^2 l_\text{SE}^2 \rho}},
\end{equation}
with Boltzmann constant $k_\text{B}$, elementary charge $e_0$, molar mass $M$, and mass density $\rho$.
Smaller values of $\lambda$ correspond to thinner \acp{scl} if identical boundary conditions are applied~\cite{Katzenmeier2021}. From this parameter, we deduce, that $l_\text{SCL} \propto \sqrt{T}$. 
\item At the low-temperature limit, i.e.~$T \to 0 \ \text{K}$, the spatial extensions $l_\text{SCL,c}$ and $l_\text{SCL,a}$ of one-dimensional \acp{scl} can be expressed analytically~\cite{BeckerSteinberger2021} as
\begin{align}
    l_\text{SCL,c} &= \sqrt{\frac{2\epsilon \Delta\Phi}{F} \frac{z_-  c_-}{(z_+  c_{\text{max}} + z_-  c_-)(z_-  c_-)  }},\\
    l_\text{SCL,a} &= -\frac{z_-}{ z_+ \frac{c_{\text{max}}}{c_-} + z_-} l_\text{c} = f_\text{sym} \cdot l_\text{c}\,.
\end{align} 
Both, $l_\text{SCL,c}$ and $l_\text{SCL,a}$ are proportional to the square root of the applied difference in electric potential~$l_\text{SCL,c},l_\text{SCL,a} \propto \sqrt{\Delta\Phi}$. Depending on the ratio $\frac{c-}{c_\text{max}}$, the respective lengths can significantly differ. Therefore, we introduce a symmetry factor $f_\text{sym}$ to quantitatively capture this asymmetry. For the material parameters used in this work, the symmetry factor is computed as~$f_\text{sym} = \frac{3}{2}$, such that $l_\text{SCL,a} > l_\text{SCL,c}$.
\end{enumerate}
Now, we can estimate the length of the \ac{scl} by the following ansatz
\begin{equation}
    l^*_\text{SCL,c,a}(\Delta \Phi, T) = (k_0 + k_1 \sqrt{T}) \cdot l_\text{SCL,c,a},
\end{equation}
with the unknown constants $k_0$ and $k_1$. By performing one-dimensional simulations with different values for $\Delta \Phi$, $T$, and blocking electrode conditions, we can compute $k_0$ and $k_1$. Additionally, we introduce a safety factor~$\zeta$ to cover the impact of complex three-dimensional microstructures which cannot be considered in this estimate and define
\begin{equation}
    \label{eq:estimate_length_SCL}
    l_\text{SCL} = \zeta \cdot \max \{l^*_c,l^*_a\}.
\end{equation}
For the aforementioned material parameters, and a safety factor of~$\zeta=2$ we get $l_\text{SCL} = 0.4 \ \mu \text{m}$. We use this value throughout the remainder of this work.\\
We verify our choice of~$l_\text{SCL}$ by considering an extreme case where the entire difference in electric potential occurs inside one \ac{scl}. Therefore, the conditions for blocking electrodes are applied, and additionally, the concentration is fixed on one side of the domain $c(\xi=l_\text{SCL}) = c_\text{bulk}$ to obtain a single-sided \ac{scl}. The result for this setup is shown in \cref{fig:validate_extension}. It can be seen, that the gradients of the concentration vanish for~$\xi > 0.2 \ \mu \text{m}$, such that $l_\text{SCL} = 0.4 \ \mu \text{m}$ is a sufficiently large choice.
\subsubsection*{Optimal discretization size of the \ac{scl} domain}
While the length of the \ac{scl} domain is obtained based on the maximal expected size of the \ac{scl}, the discretization size is determined by the minimal expected size of the \ac{scl} to resolve the change in gradients there. Thus, we perform a spatial convergence analysis by comparing the stored charge~$Q_\text{SCL}$ inside the smaller \ac{scl} for different sizes of the discretization. We conduct the simulation for $n_\text{ele} = \{40,80,160,320\}$ and choose $n_\text{ele}=2560$ as the reference solution. Thus, the reference discretization has four times the nodes of the finest discretization in the convergence study. Again, we investigate a single-sided \ac{scl}. \cref{fig:charge_error} illustrates that for the used material parameters and boundary conditions approximately~$300$ nodes are required to obtain a relative error below~$0.1\%$ which is considered as very small. We emphasize, that the required number of nodes per $l_\text{SCL}$ is a function of the symmetry factor $f_\text{sym}$. In general, higher values of $f_\text{sym}$ increase the computational effort, as it leads to a larger length~$l_\text{SCL}$ and requires a finer discretization.
\subsection*{Validation of the outlined model}
Different strategies are followed to validate the outlined model: Solving a pseudo one-dimensional problem, testing for conservation properties, and comparing the results with those obtained by a fully resolved model.
\subsubsection*{Validation of the coupled three-dimensional model as pseudo one-dimensional model}
\label{sec:pseudo_1D}
\begin{figure}[ht]
    \centering
    \def\svgwidth{8.2cm}
    \fontsize{5}{10}\selectfont
    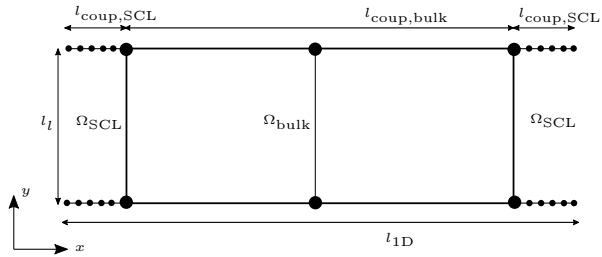
    \caption{Schematic sketch of the computational domain for comparison of the coupled three-dimensional model and the one-dimensional model.}
    \label{fig:comparison_1D_micro_macro_geometry}
\end{figure}

\begin{figure}[ht]
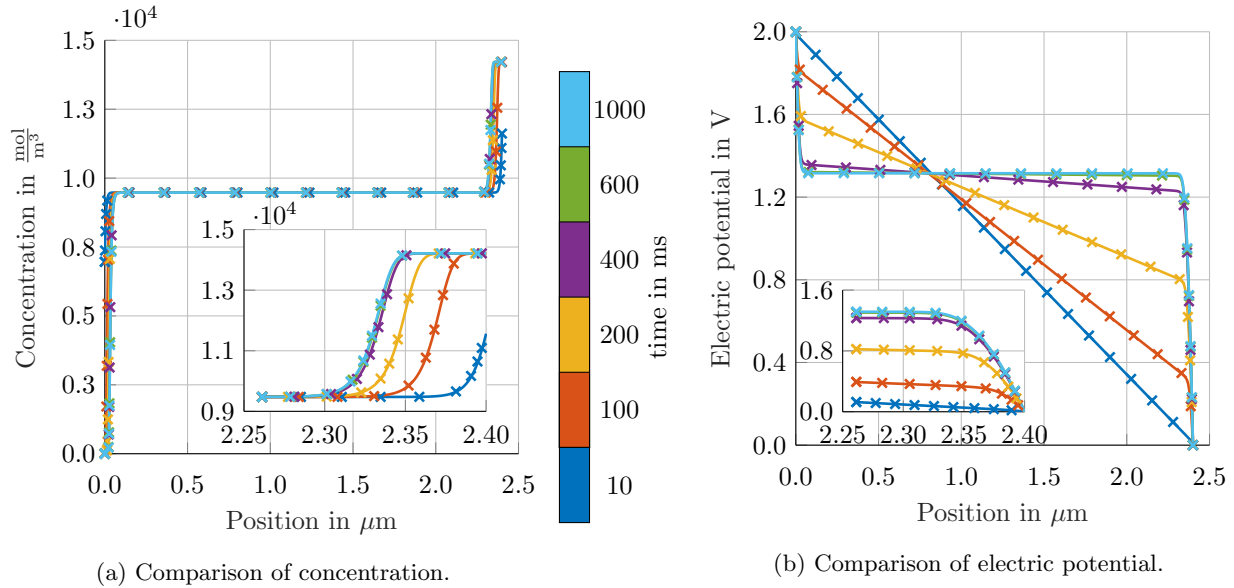

    \centering
    \begin{subfigure}{0.44\textwidth}
        \centering
        \vfill
        \input{figures/comparison_1D_micro_macro_conc.tikz}
        \caption{Comparison of concentration.}
        \label{fig:comparison_1D_micro_macro_conc}
        \vfill
    \end{subfigure}
    \hfill
    \begin{subfigure}{0.1\textwidth}
        \centering
        \vfill
        \begin{tikzpicture}

\definecolor{mycolor1}{rgb}{0.00000,0.44700,0.74100}%
\definecolor{mycolor2}{rgb}{0.85000,0.32500,0.09800}%
\definecolor{mycolor3}{rgb}{0.92900,0.69400,0.12500}%
\definecolor{mycolor4}{rgb}{0.49400,0.18400,0.55600}%
\definecolor{mycolor5}{rgb}{0.46600,0.67400,0.18800}%
\definecolor{mycolor6}{rgb}{0.30100,0.74500,0.93300}%

\filldraw[fill=mycolor6, draw=black] (0,6) rectangle (0.4,1.0);
\filldraw[fill=mycolor5, draw=black] (0,5) rectangle (0.4,1.0);
\filldraw[fill=mycolor4, draw=black] (0,4) rectangle (0.4,1.0);
\filldraw[fill=mycolor3, draw=black] (0,3) rectangle (0.4,1.0);
\filldraw[fill=mycolor2, draw=black] (0,2) rectangle (0.4,1.0);
\filldraw[fill=mycolor1, draw=black] (0,0) rectangle (0.4,1.0);

\node at (0.8,0.5) {$10$};
\node at (0.8,1.5) {$100$};
\node at (0.8,2.5) {$200$};
\node at (0.8,3.5) {$400$};
\node at (0.8,4.5) {$600$};
\node at (0.8,5.5) {$1000$};

\node at (1.3,3.0) [rotate=90] {time in ms};

\end{tikzpicture}
        \vfill
    \end{subfigure}
    \hfill
    \begin{subfigure}{0.44\textwidth}
        \centering
        \vfill
        \input{figures/comparison_1D_micro_macro_pot.tikz}
        \caption{Comparison of electric potential.}
        \label{fig:comparison_1D_micro_macro_pot}
        \vfill      
    \end{subfigure}
    \caption{Comparison of concentration and electric potential for the pure \ac{scl} model (crosses) and the coupled \ac{scl} model (solid lines). The different time steps are assigned to the lines by color codes. The small figures inside the plots represent zooms into the \ac{scl} domains at the right side with a finer resolution of the crosses. Note, that the crosses do not represent the spatial discretization.}
\end{figure}

\begin{figure}[ht]
    \centering
    \begin{subfigure}{0.85\textwidth}
        \centering
        \vfill
        \input{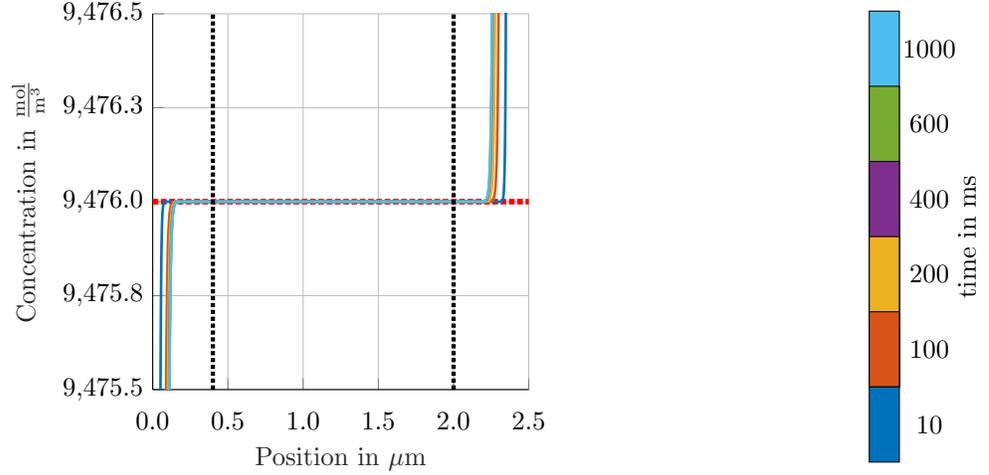}
        \vfill
    \end{subfigure}
    \hfill
    \begin{subfigure}{0.1\textwidth}
        \centering
        \vfill
        \begin{tikzpicture}

\definecolor{mycolor1}{rgb}{0.00000,0.44700,0.74100}%
\definecolor{mycolor2}{rgb}{0.85000,0.32500,0.09800}%
\definecolor{mycolor3}{rgb}{0.92900,0.69400,0.12500}%
\definecolor{mycolor4}{rgb}{0.49400,0.18400,0.55600}%
\definecolor{mycolor5}{rgb}{0.46600,0.67400,0.18800}%
\definecolor{mycolor6}{rgb}{0.30100,0.74500,0.93300}%

\filldraw[fill=mycolor6, draw=black] (0,6) rectangle (0.4,1.0);
\filldraw[fill=mycolor5, draw=black] (0,5) rectangle (0.4,1.0);
\filldraw[fill=mycolor4, draw=black] (0,4) rectangle (0.4,1.0);
\filldraw[fill=mycolor3, draw=black] (0,3) rectangle (0.4,1.0);
\filldraw[fill=mycolor2, draw=black] (0,2) rectangle (0.4,1.0);
\filldraw[fill=mycolor1, draw=black] (0,0) rectangle (0.4,1.0);

\node at (0.8,0.5) {$10$};
\node at (0.8,1.5) {$100$};
\node at (0.8,2.5) {$200$};
\node at (0.8,3.5) {$400$};
\node at (0.8,4.5) {$600$};
\node at (0.8,5.5) {$1000$};

\node at (1.3,3.0) [rotate=90] {time in ms};

\end{tikzpicture}
        \vfill
    \end{subfigure}
    \caption{Zoom to concentration in the bulk domain computed with the pure one-dimensional model at different time steps approximately equals the bulk concentration (red line) inside the bulk domain. The vertical lines separate the bulk domain from the \ac{scl} domain.}
    \label{fig:comparison_1D_micro_macro_bulk_concentration}
\end{figure}

\begin{figure}[H]
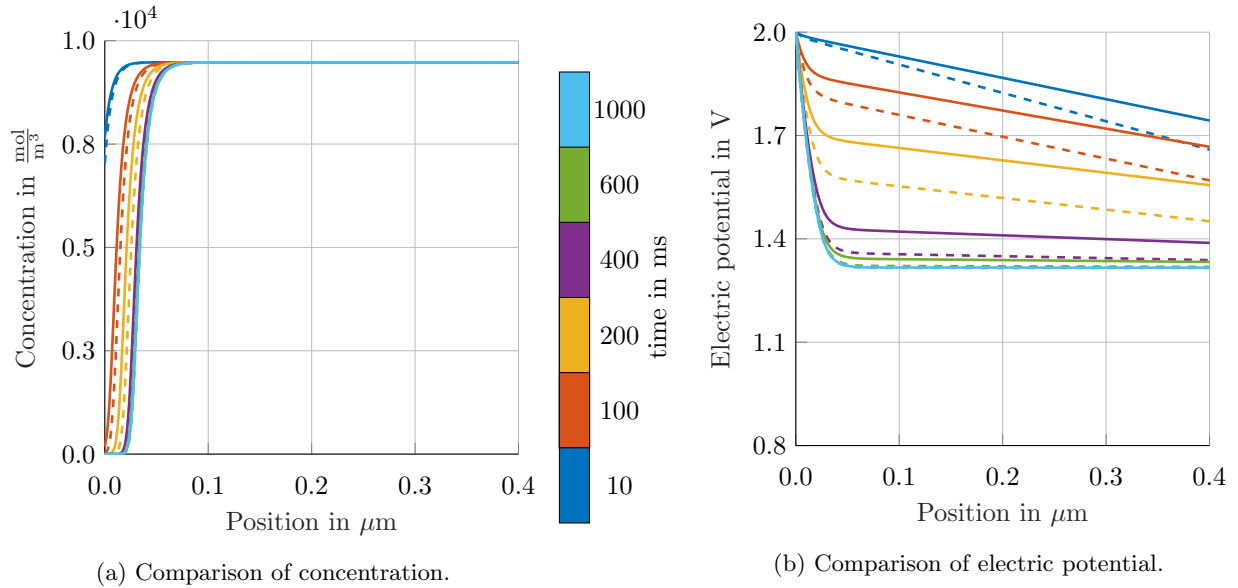

    \centering
    \begin{subfigure}{0.44\textwidth}
        \centering
        \vfill
        \input{figures/comparison_matching_enlarged_zoom_conc.tikz}
        \caption{Comparison of concentration.}
        \label{fig:comparison_matching_enlarged_zoom_conc}
        \vfill
    \end{subfigure}
    \hfill
    \begin{subfigure}{0.1\textwidth}
        \centering
        \vfill
        \begin{tikzpicture}

\definecolor{mycolor1}{rgb}{0.00000,0.44700,0.74100}%
\definecolor{mycolor2}{rgb}{0.85000,0.32500,0.09800}%
\definecolor{mycolor3}{rgb}{0.92900,0.69400,0.12500}%
\definecolor{mycolor4}{rgb}{0.49400,0.18400,0.55600}%
\definecolor{mycolor5}{rgb}{0.46600,0.67400,0.18800}%
\definecolor{mycolor6}{rgb}{0.30100,0.74500,0.93300}%

\filldraw[fill=mycolor6, draw=black] (0,6) rectangle (0.4,1.0);
\filldraw[fill=mycolor5, draw=black] (0,5) rectangle (0.4,1.0);
\filldraw[fill=mycolor4, draw=black] (0,4) rectangle (0.4,1.0);
\filldraw[fill=mycolor3, draw=black] (0,3) rectangle (0.4,1.0);
\filldraw[fill=mycolor2, draw=black] (0,2) rectangle (0.4,1.0);
\filldraw[fill=mycolor1, draw=black] (0,0) rectangle (0.4,1.0);

\node at (0.8,0.5) {$10$};
\node at (0.8,1.5) {$100$};
\node at (0.8,2.5) {$200$};
\node at (0.8,3.5) {$400$};
\node at (0.8,4.5) {$600$};
\node at (0.8,5.5) {$1000$};

\node at (1.3,3.0) [rotate=90] {time in ms};

\end{tikzpicture}
        \vfill
    \end{subfigure}
    \hfill
    \begin{subfigure}{0.44\textwidth}
        \centering
        \vfill
        \input{figures/comparison_matching_enlarged_zoom_pot.tikz}
        \caption{Comparison of electric potential.}
        \label{fig:comparison_matching_enlarged_zoom_pot}
        \vfill      
    \end{subfigure}
    \caption{Comparison of concentration and electric potential for the matching geometry (dashed lines) and the enlarged geometry (solid lines). The different time steps are assigned to the lines by color codes.}
    \label{fig:comparison_matching_enlarged}
\end{figure}

We compare the results of our coupled model that combines one- and three-dimensional discretizations ("coupled \ac{scl} model") with the result of a pure one-dimensional model as shown before ("pure \ac{scl} model"). For comparison, the x-dimension of the coupled approach matches exactly the length of the pure one-dimensional model $l_\text{coup} = l_{1D}$. The length of the coupled problem is $l_\text{coup} = 2 \ l_\text{coup,SCL} + l_\text{coup,bulk}$ (see \cref{fig:comparison_1D_micro_macro_geometry}).
The other two dimensions in the coupled \ac{scl} model are chosen such that a reasonable aspect ratio of the three-dimensional domain is maintained. The three-dimensional domain of the coupled \ac{scl} model is discretized using two equal-sized hexahedral elements with linear shape functions, while the one-dimensional domain of the coupled \ac{scl} model consists of $2400$ line elements with linear shape functions, meaning $300$ nodes in each \ac{scl} discretization. The pure \ac{scl} model is discretized with $1800$ line elements, such that the discretization inside the \ac{scl} domain is identical for both models. \\
We prescribe a scenario with blocking electrodes: A difference in potential $\Delta \Phi$ is applied to both ends of the domain, while the flux of mass outside of the considered domain is prohibited. All relevant parameters are summarized in \cref{table:params_comparison_1D_micro_macro}.
\begin{table}[ht]
    \centering
    \caption{Parameters for comparing results between pure and coupled three-dimensional Space-Charge-Layer model.}
    \begin{tabular}{c | c | c}
      \textbf{quantity}                   & \textbf{symbol}          & \textbf{value} \\
      \hline
      length of domain                    & $l_\text{coup} = l_{1D}$ & $2.4 \ \mu \text{m}$ \\
      lateral length                      & $l_\text{l}$             & $0.4 \ \mu \text{m}$ \\
      length of SCL domain                & $l_\text{coup,SCL}$      & $0.4 \ \mu \text{m}$ \\
      \hline
      coupled: number of elements (SCL)   & $n_\text{ele, SCL}$      & $2400$ \\
      coupled: number of elements (bulk)  & $n_\text{ele, SCL}$      & $2$ \\
      pure: number of elements            & $n_\text{ele, 1D}$       & $1800$ \\
      time step size                      & $\Delta t$               & $1 \ \text{ms}$ \\
      \hline
      total time                          & $t_\text{max}$           & $1 \ \text{s}$
    \end{tabular}
    \label{table:params_comparison_1D_micro_macro}
\end{table}
In \cref{fig:comparison_1D_micro_macro_conc} we compare the results of the concentration and in \cref{fig:comparison_1D_micro_macro_pot} the electric potential from the pure \ac{scl} model with the results from the coupled \ac{scl} model for different points in time.
The results computed with the two models are in very good agreement. Even in regions where the curvature of both the concentration and the electric potential, changes most (see zooms) the deviation is negligible.\\
Analyzing the results of the pure one-dimensional model allows quantifying the approximation error labeled as "model error". In \cref{fig:comparison_1D_micro_macro_bulk_concentration}, it can be seen that the concentration inside the bulk domain is very close to the bulk concentration $c_\text{bulk}$ throughout the entire simulation time.
 \noindent
This shows that the only assumption in the derivation for the equations of the bulk domain, namely that the concentration remains at the fixed value $c_\text{bulk}$, is justified. \\
Besides, we want to quantify the approximation error, which we labeled "geometric error" by modifying the dimensions of~$\Omega_\text{bulk}$. As shown in \cref{fig:comparison_1D_micro_macro_pot}, most of the potential drop occurs inside the \acp{scl} except for the first instances of time. This already shows that the influence of a slightly larger domain has a negligible influence on the global shape of the potential and the concentration. To investigate this in more detail, we choose~$l_\text{coup,bulk} = l_{1D}$ and keep the size of the \ac{scl} domain untouched, such that $l_\text{coup} > l_{1D}$. In \cref{fig:comparison_matching_enlarged} we compare the results of the matching geometric size with the results of the enlarged geometry by zooming into the plot of the concentration (\cref{fig:comparison_matching_enlarged_zoom_conc}) and the electric potential (\cref{fig:comparison_matching_enlarged_zoom_pot}) at the \ac{scl} on the left side of the domain.
 \noindent
As expected, we introduce an error in the shape of the electric potential that decreases towards the steady state, while the error in the concentration in the bulk is negligible. However, in this academic example, we triggered the geometric error on purpose to show its influence but want to emphasize, that here $l_\text{SCL} << l_\text{SE}$ is not valid anymore.
\subsubsection*{Validation of conservation properties}
\begin{figure}[ht]
    \begin{floatrow}
    \ffigbox{
    \def\svgwidth{0.4\textwidth}
    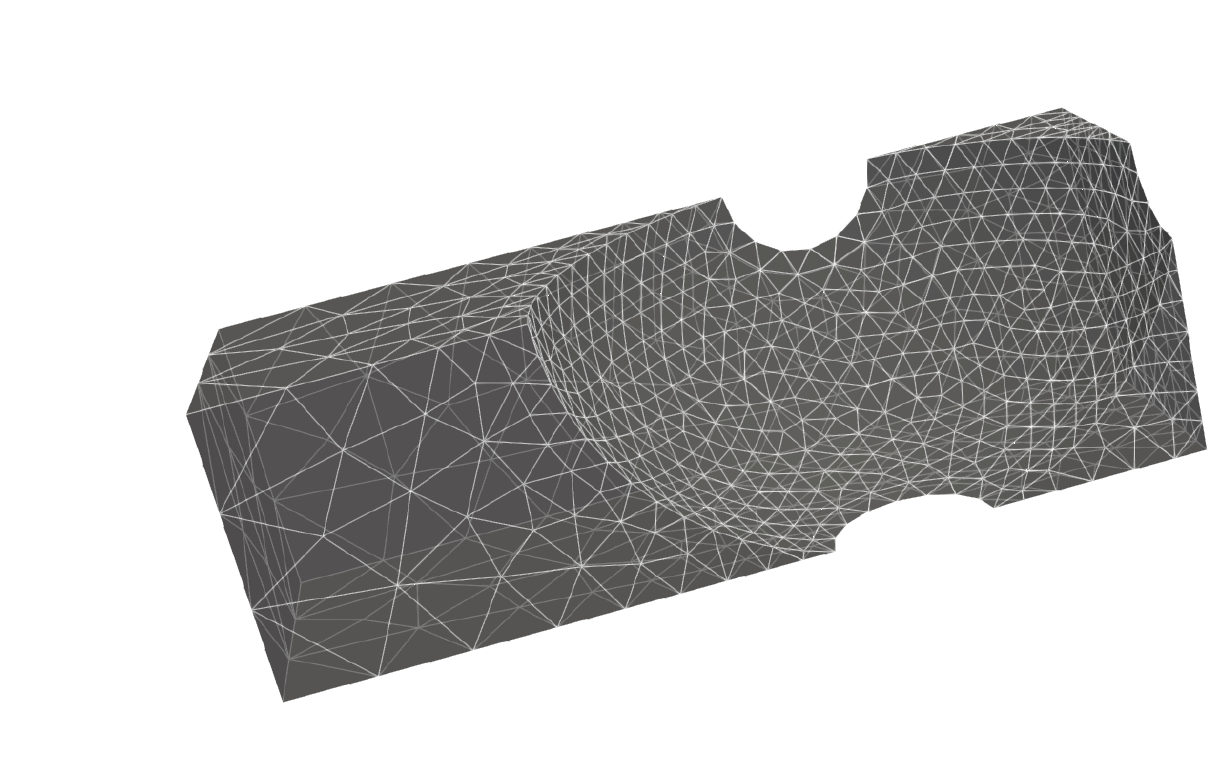
    }{
      \caption{Geometric representation of the geometry for validation of conservation properties.}
      \label{fig:1_sphere_geometry}
    }
    \capbtabbox{
    \begin{tabular}{c | c | c}
    \textbf{quantity}    & \textbf{symbol}     & \textbf{value} \\
    \hline
    length of domain     & $l$                 & $3 \ \mu\text{m}$ \\
    diameter of sphere   & $d$                 & $2 \ \mu\text{m}$ \\
    lateral length       & $l_\text{l}$        & $0.97 \ \mu\text{m}$ \\
    length separator     & $l_\text{s}$        & $1 \ \mu\text{m}$ \\
    length of SCL domain & $l_\text{coup,SCL}$ & $0.4 \ \mu\text{m}$ \\
    \hline
    number of nodes bulk & $n_\text{bulk}$     & $2,384$ \\
    number of nodes SCL  & $n_\text{SCL}$      & $103,500$ \\
    time step size       & $\Delta t$          & $5 \ \text{ms}$ \\
    \hline
    total time           & $t_\text{max}$      & $10 \ \text{s}$
\end{tabular}
    }{
      \caption{Parameters for testing conservation properties.}
      \label{table:params_test_conservation_properties}
    }
    \end{floatrow}
\end{figure}

\begin{figure}[H]
    \centering
    \def\svgwidth{7.0cm}
\begingroup%
  \makeatletter%
  \providecommand\color[2][]{%
    \errmessage{(Inkscape) Color is used for the text in Inkscape, but the package 'color.sty' is not loaded}%
    \renewcommand\color[2][]{}%
  }%
  \providecommand\transparent[1]{%
    \errmessage{(Inkscape) Transparency is used (non-zero) for the text in Inkscape, but the package 'transparent.sty' is not loaded}%
    \renewcommand\transparent[1]{}%
  }%
  \providecommand\rotatebox[2]{#2}%
  \newcommand*\fsize{\dimexpr\f@size pt\relax}%
  \newcommand*\lineheight[1]{\fontsize{\fsize}{#1\fsize}\selectfont}%
  \ifx\svgwidth\undefined%
    \setlength{\unitlength}{1038.75bp}%
    \ifx\svgscale\undefined%
      \relax%
    \else%
      \setlength{\unitlength}{\unitlength * \real{\svgscale}}%
    \fi%
  \else%
    \setlength{\unitlength}{\svgwidth}%
  \fi%
  \global\let\svgwidth\undefined%
  \global\let\svgscale\undefined%
  \makeatother%
  \begin{picture}(1,0.97515332)%
    \lineheight{1}%
    \setlength\tabcolsep{0pt}%
    \put(0,0){\includegraphics[width=\unitlength,page=1]{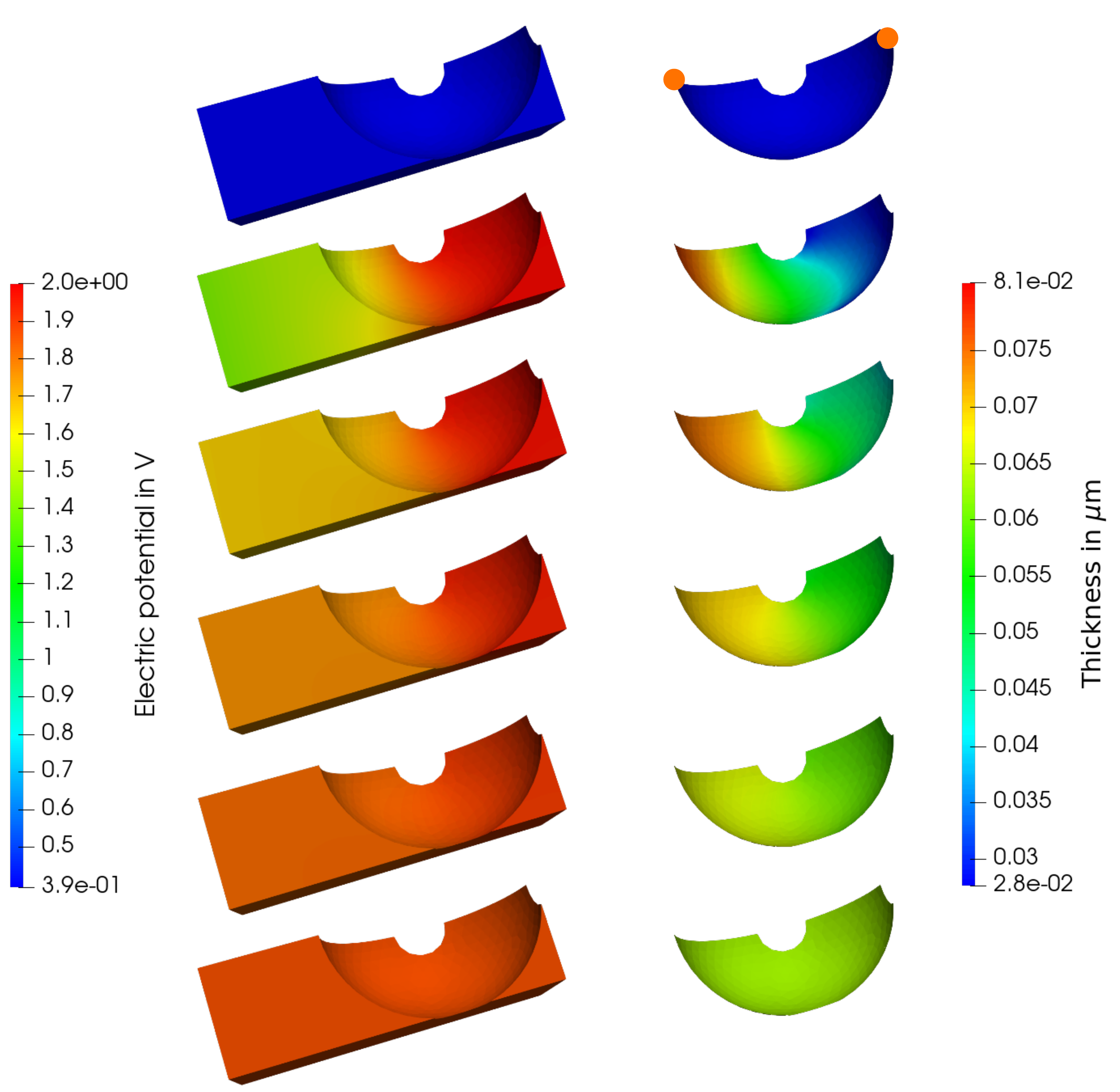}}%
    \put(0.53407692,0.92742428){\color[rgb]{0,0,0}\makebox(0,0)[lt]{\lineheight{1.25}\smash{\begin{tabular}[t]{l}$d_\text{SCL,c,max}$\end{tabular}}}}%
    \put(0.74385612,0.96052487){\color[rgb]{0,0,0}\makebox(0,0)[lt]{\lineheight{1.25}\smash{\begin{tabular}[t]{l}$d_\text{SCL,c,min}$\end{tabular}}}}%
  \end{picture}%
\endgroup%

    \caption{Temporal development of electric potential and thickness of \ac{scl} from the initial state to the steady state (top to bottom). The locations where the minimal and maximal thickness of the \ac{scl} occurs are highlighted with an orange dot.}
    \label{fig:1_sphere_results}
\end{figure}

\begin{figure}[ht]
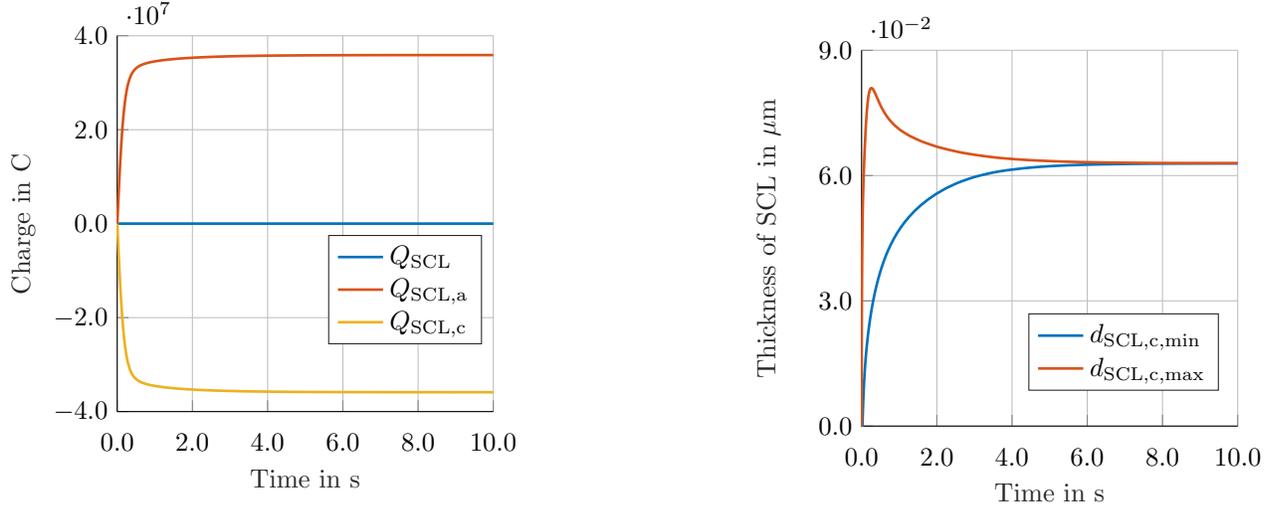

    \begin{subfigure}{0.4\textwidth}
        \centering
        \vfill
        \input{figures/1_sphere_conservation.tikz}     
        \caption{Temporal development of the integrated deviation of the charge from the charge-neutral state for the \ac{scl} at the anode, at the cathode, and their sum.}
        \label{fig:1_sphere_conservation}
        \vfill
    \end{subfigure}
    \hfill
    \begin{subfigure}{0.4\textwidth}
        \centering
        \vfill
        \input{figures/1_sphere_min_max_thickness.tikz}
        \caption{Temporal development of minimal and maximal thickness of \acp{scl} at the spherical electrode.}
        \label{fig:1_sphere_min_max_thickness}
        \vfill
    \end{subfigure}
    \caption{Results for the validation of conservation properties.}
\end{figure}

For the validation of conservation properties, we use a geometry, that is not pseudo-one-dimensional but still as simple as possible (see \cref{fig:1_sphere_geometry}).
It consists of one spherical particle embedded into the bulk solid electrolyte representing the cathode (e.g.~NMC). The anode is represented by a planar surface (e.g.~lithium metal). The size of the geometry is reduced by making use of symmetry. All relevant parameters are listed in \cref{table:params_test_conservation_properties}. Again, we apply a difference in potential $\Delta \Phi$ without any flux of mass between both electrodes to represent blocking electrodes and observe the transient behavior until the steady state is reached. For visualization, we take snapshots at $t=\{0, 0.31, 0.625, 1.25, 2.5, 10.0 \} \ \text{s}$ to present the three-dimensionally resolved development of the electric potential and of the thickness of the \ac{scl} at $\Gamma_\text{SCL-c}$ over time (see \cref{fig:1_sphere_results}).
It is clearly visible, that the thickness of the \ac{scl} changes over time and also significantly varies at different spatial positions. As expected, the electric potential converges towards a constant value in the steady state and thus, also the thickness converges towards a constant value as the thickness is determined by the difference in potential across the \ac{scl}.\\
We expect the integrated deviation of charge from the neutrally charged state as defined before to remain constant over time due to the conservation of mass. For visualization, we split this integral into one part at the anode $Q_\text{SCL,a}$ and one part at the cathode $Q_\text{SCL,c}$. In \cref{fig:1_sphere_conservation} we show the development of $Q_\text{SCL}$, $Q_\text{SCL,a}$, and $Q_\text{SCL,c}$ over time.
We can clearly see that $Q_\text{SCL,a}$ increases, while $Q_\text{SCL,c}$ decreases over time, but their sum remains - in the expected bounds of the numerical accuracy - constant. This shows that our formulation guarantees conservation properties.\\ 
From this geometrically simple example, we can derive further insights beyond the proof of conservation properties considering the development of the thickness of the \ac{scl}, which strongly differs depending on its local position. While the thickness of the \ac{scl} on $\Gamma_\text{SCL-c}$ close to the anode develops instantly, the \ac{scl} on the opposite side on $\Gamma_\text{SCL-c}$ develops much slower. In \cref{fig:1_sphere_min_max_thickness} we compare the minimal thickness $d_\text{SCL,c,min} = \text{min}(d_\text{SCL,c})$ and the maximal thickness $d_\text{SCL,c,max} = \text{max}(d_\text{SCL,c,max})$ of the \ac{scl} at the interface to the cathode. The minimal thickness converges monotonically towards the steady state, while the maximal thickness rapidly increases to a value, that is larger than the final value at the steady state and eventually decreases again towards the thickness at the steady state. This unintuitive behavior can be explained by investigating the impedance between both electrodes (see \cref{fig:1_sphere_results}). The total impedance is composed of the sum of the impedance in the bulk and in the \ac{scl} and its minimal value defines the favored conduction path. The impedance in the bulk electrolyte scales with the length through the electrolyte. The impedance in the \ac{scl} increases with increased stored charge. During the transient phase, the minimal total impedance continuously changes, as the impedance from the \ac{scl} changes due to more stored charge. Thus, the favored conduction path changes to regions with more contributions from the impedance of the bulk. This can be observed in the electric potential inside the bulk electrolyte which at the beginning features a gradient only between the anode and regions on~$\Gamma_\text{SCL-c}$ closest to the anode. Later, the gradient is visible inside the entire bulk domain, before it vanishes completely in the steady state.\\
Additionally, we observe in the steady state that the electric potential in the bulk domain differs from the electric potential in the pseudo one-dimensional case computed in the examples before. This is caused by the different areas of the interfaces with the anode and the cathode, respectively. As shown before, the total charge within both \acp{scl} sum up to zero, but due to the different interface areas, the local charge density is different, and thus, the entire shape of the \ac{scl}. Again, this highlights the necessity to three-dimensionally resolve \acp{scl}.
\subsubsection*{Comparison of the solution without simplification assumptions}
\begin{figure}[ht]
    \begin{floatrow}
    \ffigbox{
    \def\svgwidth{0.3\textwidth}
\begingroup%
  \makeatletter%
  \providecommand\color[2][]{%
    \errmessage{(Inkscape) Color is used for the text in Inkscape, but the package 'color.sty' is not loaded}%
    \renewcommand\color[2][]{}%
  }%
  \providecommand\transparent[1]{%
    \errmessage{(Inkscape) Transparency is used (non-zero) for the text in Inkscape, but the package 'transparent.sty' is not loaded}%
    \renewcommand\transparent[1]{}%
  }%
  \providecommand\rotatebox[2]{#2}%
  \newcommand*\fsize{\dimexpr\f@size pt\relax}%
  \newcommand*\lineheight[1]{\fontsize{\fsize}{#1\fsize}\selectfont}%
  \ifx\svgwidth\undefined%
    \setlength{\unitlength}{700.36964997bp}%
    \ifx\svgscale\undefined%
      \relax%
    \else%
      \setlength{\unitlength}{\unitlength * \real{\svgscale}}%
    \fi%
  \else%
    \setlength{\unitlength}{\svgwidth}%
  \fi%
  \global\let\svgwidth\undefined%
  \global\let\svgscale\undefined%
  \makeatother%
  \begin{picture}(1,0.98680814)%
    \lineheight{1}%
    \setlength\tabcolsep{0pt}%
    \put(0,0){\includegraphics[width=\unitlength,page=1]{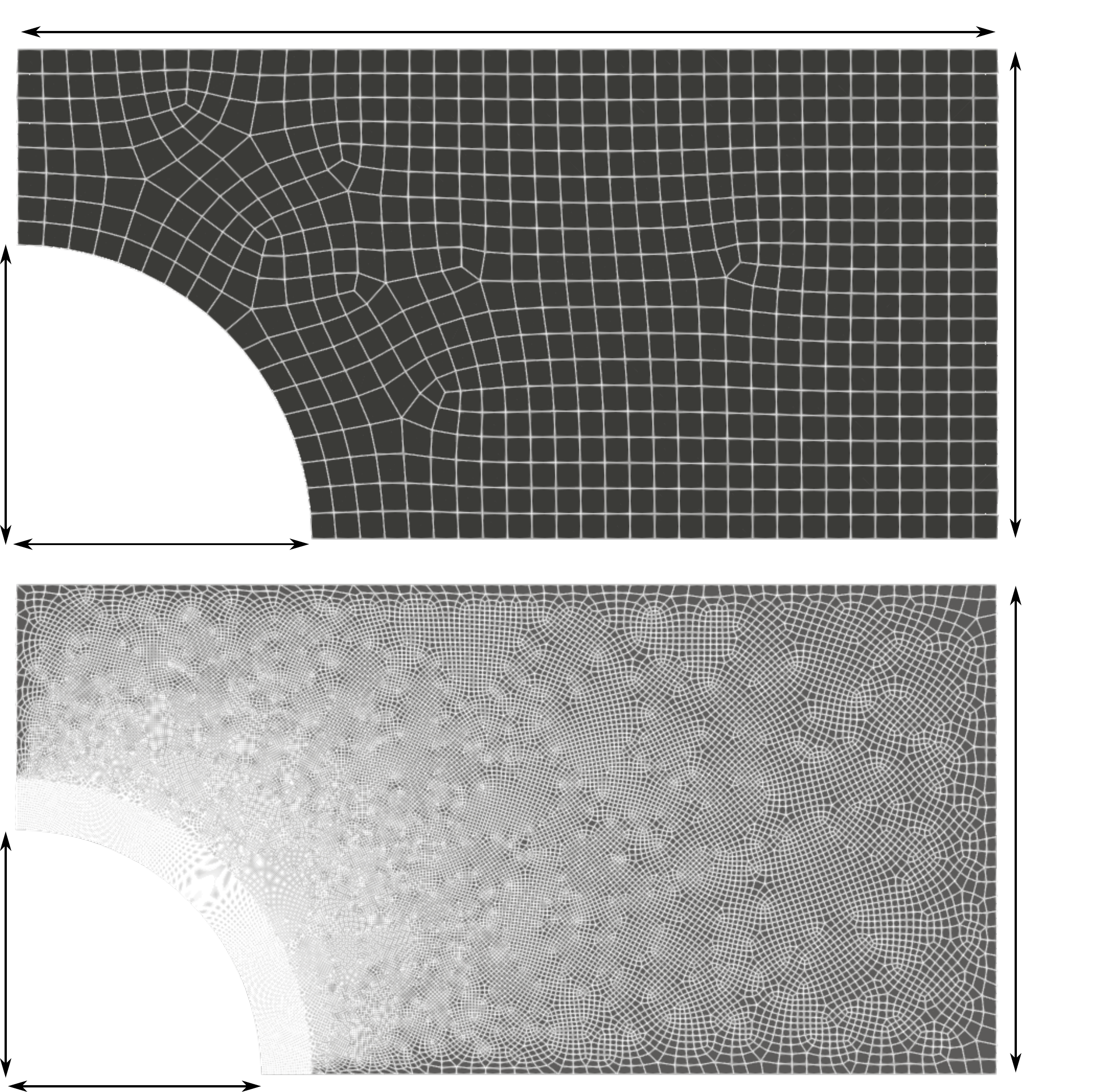}}%
    \put(0.11367339,0.03604265){\color[rgb]{0,0,0}\makebox(0,0)[lt]{\lineheight{1.25}\smash{\begin{tabular}[t]{l}$r$\end{tabular}}}}%
    \put(0.01811382,0.10833578){\color[rgb]{0,0,0}\makebox(0,0)[lt]{\lineheight{1.25}\smash{\begin{tabular}[t]{l}$r$\end{tabular}}}}%
    \put(0.12083446,0.51967442){\color[rgb]{0,0,0}\makebox(0,0)[lt]{\lineheight{1.25}\smash{\begin{tabular}[t]{l}$r_\text{c}$\end{tabular}}}}%
    \put(0.01859906,0.61627923){\color[rgb]{0,0,0}\makebox(0,0)[lt]{\lineheight{1.25}\smash{\begin{tabular}[t]{l}$r_\text{c}$\end{tabular}}}}%
    \put(0.92734379,0.70419075){\color[rgb]{0,0,0}\makebox(0,0)[lt]{\lineheight{1.25}\smash{\begin{tabular}[t]{l}$l_\text{v}$\end{tabular}}}}%
    \put(0.92984173,0.2241986){\color[rgb]{0,0,0}\makebox(0,0)[lt]{\lineheight{1.25}\smash{\begin{tabular}[t]{l}$l_\text{v}$\end{tabular}}}}%
    \put(0.44349852,0.97812968){\color[rgb]{0,0,0}\makebox(0,0)[lt]{\lineheight{1.25}\smash{\begin{tabular}[t]{l}$l_\text{h}$\end{tabular}}}}%
  \end{picture}%
\endgroup%

    }{
      \caption{Geometric representation of the coupled model (top) and the model without assumptions (bottom).}
      \label{fig:coupled_pure_geometry}
    }
    \capbtabbox{
    \begin{tabular}{P{4cm} | P{1.2cm }| P{1cm}}
    \textbf{quantity}    & \textbf{symbol}     & \textbf{value} \\
    \hline
    radius of cylinder               & $r         $   & $0.5 \ \mu \text{m}$  \\
    radius of cylinder (coupled)     & $r_\text{c}$   & $0.6 \ \mu \text{m}$ \\
    vertical edge length             & $l_\text{v}$   & $1 \ \mu \text{m}$ \\
    horizontal edge length           & $l_\text{h}$   & $2 \ \mu \text{m}$ \\
    length of SCL domain (coupled)   & $l_\text{SCL}$ & $0.1 \ \mu \text{m}$ \\
    \hline
    number of nodes                  & $n$            & $150650$ \\
    number of nodes (coupled, total) & $n_\text{c}$   & $13136$ \\
    time step size                   & $\Delta t$     &  $0.2 \ \text{s}$  \\
    \hline
    total time                       & $t_\text{max}$ & $200 \ \text{s}$
\end{tabular}
    }{
      \caption{Parameters for comparison of the coupled model and the model without assumptions.}
      \label{table:params_coupled_pure}
    }
    \end{floatrow}
\end{figure}

\begin{figure}[ht]
    \centering
    \begin{subfigure}{8.2cm}
        \centering
        \vfill
        \includegraphics[width = \textwidth]{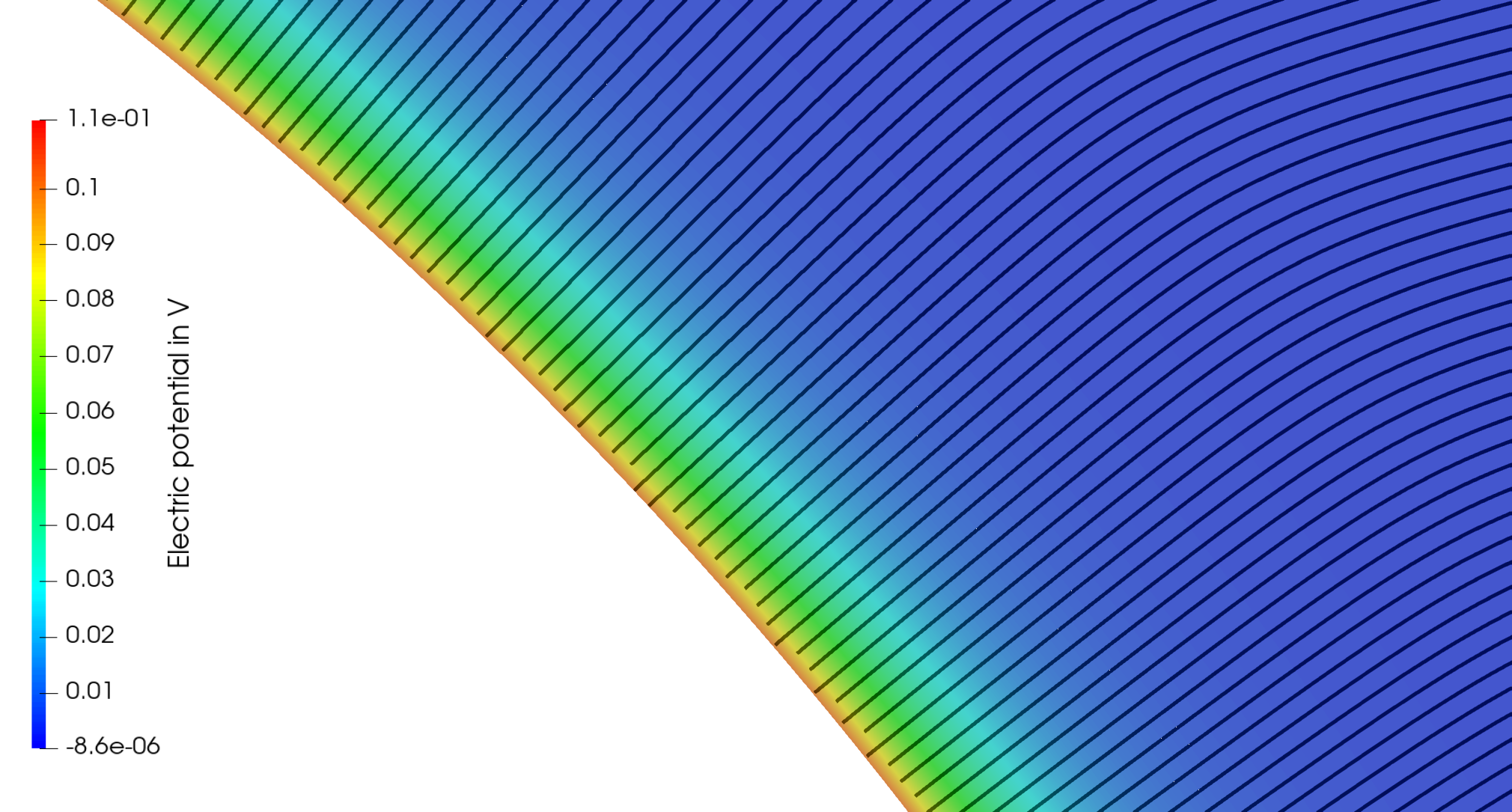}
        \caption{Electric potential (color bar) and gradient of electric potential (streamlines) in the region of the \ac{scl} in the steady state.}
        \label{fig:comparison_pure_coupled_gradient}
        \vfill
    \end{subfigure}
    \hfill
    \begin{subfigure}{0.4\textwidth}
        \centering
        \vfill
        \input{figures/comparison_coupled_pure.tikz}
        \caption{Temporal development of the maximal and minimal thickness of the \ac{scl} computed with the model without assumptions (blue) and the coupled model (red).}
        \label{fig:comparison_pure_coupled_thickness}
        \vfill      
    \end{subfigure}
    \caption{Results to quantify the compatibility error.}
\end{figure}

For further validation of the proposed model, we compare the solution computed with the proposed model with the solution computed with a model without further assumptions, i.e.~solving the non-reduced equations in all three dimensions of space. The geometry for both models is shown in \cref{fig:coupled_pure_geometry} and the respective parameters are summarized in \cref{table:params_coupled_pure}.
The geometric dimensions are chosen such that the domain of the coupled model including both the bulk domain and the \ac{scl} domain equals the domain of the model without assumptions. The electrode is represented by a cylinder. To reduce the computational effort of the models, we design the problem as two-dimensional, with constant thickness in the third dimension of space. Additionally, we set the concentration at the right boundary to $c_\text{bulk}$ to obtain a single-sided \ac{scl} at the cylindrical electrode. Thus, we need a strong refinement of the mesh only at the cylindrical electrode. In contrast to the examples before, we choose the difference in electric potential to $\Delta \Phi = 0.1 \ \text{V}$ to reduce the size of the \ac{scl}, which is evaluated according to \cref{eq:estimate_length_SCL} to $l_\text{SCL} \approx 0.1 \ \mu \text{m}$. The difference in electric potential is applied between the cylindrical electrode and the right boundary.\\
First, we want to justify the assumption, that the main effects inside the \ac{scl} are one-dimensional and thus estimate the compatibility error. Therefore, we visualize the gradient of the electric potential computed with the model without assumptions (see \cref{fig:comparison_pure_coupled_gradient}).
It can be seen that the gradient at the interface to the electrode is perfectly perpendicular as expected. With increasing distance to the electrode, the direction of the gradient begins to deviate from the perpendicular direction. This means that the assumption, that the effects inside the \ac{scl} are mainly one-dimensional is sufficiently satisfied, as long as the domain $\Omega_\text{SCL}$ is small enough.\\
By comparing the temporal development of the minimal and maximal thickness of the \ac{scl} between the model without assumptions and the coupled model, we observe a good agreement. As expected, the deviation of the minimal thickness is larger compared to the maximal thickness. The minimal thickness occurs at the left-most point on the cylindrical electrode where the normal vector is perpendicular to the main direction of the gradient of the electric potential inside the bulk domain, as cations cannot redistribute tangentially within the \ac{scl}. Instead, they need to travel through the bulk domain in order to move tangentially to the surface before entering another \ac{scl} domain. Therefore, the redistribution paths are longer compared to those in the fully resolved model. Thus, the compatibility error is more prominent during the equilibration process than in the steady state. There, nearly perfect alignment of $d_\text{SCL}$ can be observed between both models, as all tangential redistribution is accomplished.\\
This model allows not just the comparison of physically meaningful quantities but also to compare the differences in computational efficiency. While the CPU time of the model without assumptions was in the order of days, the coupled model was solved within minutes.
\subsection*{Numerical experiment using a realistic microstructure}
Beyond the academic examples we showed before to validate the proposed approach, we want to apply the model to a geometrically realistic microstructure to show its capabilities.
\subsubsection*{Geometric representation and spatial discretization}
\begin{figure}[ht]
    \centering
    \def\svgwidth{0.4\textwidth}
    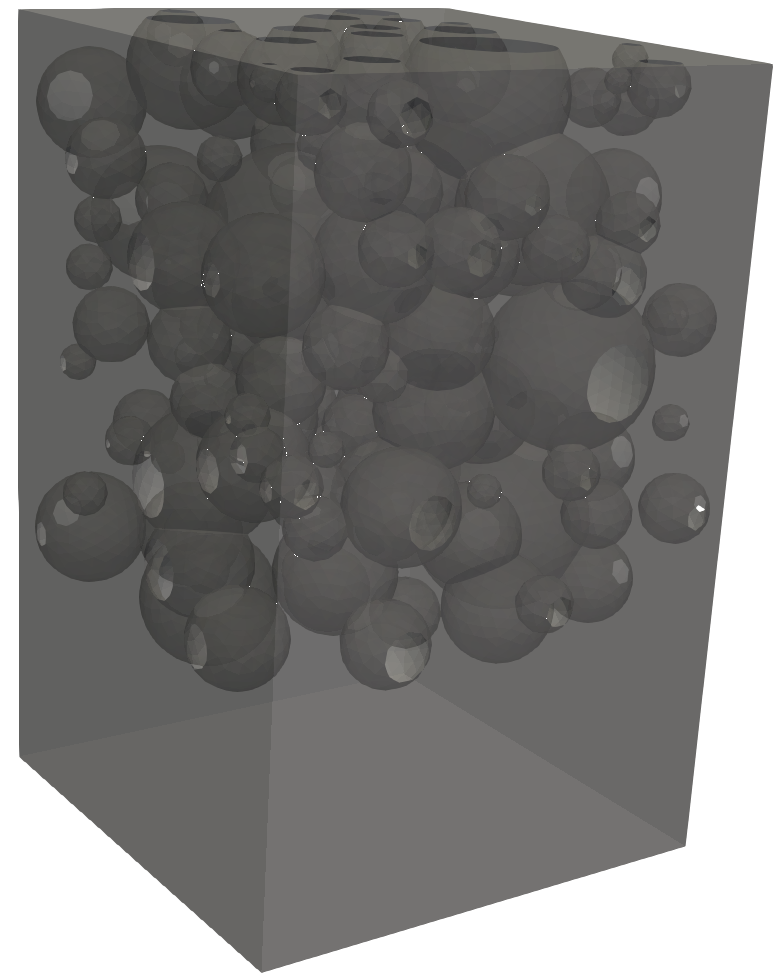
    \caption{Dimensions of the realistic geometry.}
    \label{fig:realistic_geometry}
\end{figure}

We create the geometric representation of the realistic microstructure using a setup, where perfectly shaped spherical particles as the active material of the cathode (e.g.~NMC particles), and a planar foil as the anode (e.g.~lithium metal) are assumed (see~\cref{fig:realistic_geometry} and~\cref{table:geometric_params_realisitc_geometry}).
\begin{table}[ht]
  \centering
  \caption{Parameters for the simulation of the realistic geometry.}
  \begin{tabular}{ P{9cm} | P{1.5cm }| P{3.5cm} }
    \textbf{quantity}                          & \textbf{symbol}   & \textbf{value} \\
    \hline
    length of domain                                                               & $l$                                                & $70 \ \mu \text{m}$ \\
    length of solid electrolyte separator                                          & $l_\text{s}$                                       & $17 \ \mu \text{m}$ \\
    lateral length                                                                 & $l_\text{l}$                                       & $45 \ \mu \text{m}$ \\
    log-normal distribution of diameter of cathode particles                       & \begin{tabular}{c} $\mu$ \\ $\sigma$ \end{tabular} & \begin{tabular}{c} $1.8189$ \\ $0.4589$ \end{tabular} \\
    volumetric ratio of AM and SE in composite cathode                             & $r$                                                & $0.502$ \\
    length of SCL domain                                                           & $l_\text{SCL}$                                     & $0.15 \ \mu \text{m}$ \\
    \hline
    number of nodes in bulk domain                                                 & $n_\text{bulk}$                                    & $130,211$ \\
    number of nodes in SCL domain                                                  & $n_\text{SCL}$                                     & $5,998,200$\\
    size of time step                                                              & $\Delta t$                                         & $\left\{
      \begin{array}{ll}
          50 \ \text{ms} & \, \text{if} \ t < 5 \ \text{s} \\
          300 \ \text{ms} & \, \textrm{else} \\
      \end{array}
      \right.$ \\
    \hline
    total time                                                                     & $t_\text{max}$                                     & $500 \ \text{s}$
  \end{tabular}
  \label{table:geometric_params_realisitc_geometry}
\end{table}
For this purpose we employed the following workflow: The domain of the solid electrolyte is split into the separator $\Omega_\text{bulk,s}$ and the part of the solid electrolyte inside the composite cathode $\Omega_\text{bulk, CC}$. Both domains have the same lateral length $l_\text{l}$, while their axial length $l_\text{s}$, and $l_\text{CC} = l - l_\text{s}$ differs. The diameter of the cathode particles follows a log-normal distribution with mean~$\mu$ and variance~$\sigma^2$. We create as many particles following the log-normal distribution as needed to satisfy a given volumetric ratio~$r$ of the active material and the solid electrolyte. The position of the center points of the particles is computed using a simulation with the discrete element method to obtain a spatially realistic distribution of the particles. Consequently, the \ac{scl} domain is defined on that surface $\Omega_\text{SCL} = \Gamma_\text{SCL-bulk} \times l_\text{SCL}$. The interface of the solid electrolyte and the cathode is on the surface of the spheres, and the interface of the solid electrolyte and the anode is the planar surface at the bottom of \cref{fig:realistic_geometry}. We discretize the geometry using tetrahedral elements.  The interface nodes of the bulk solid electrolyte domain and of the electrode domain are connected with line elements representing the \ac{scl} domain.
\subsubsection*{Results}
\begin{figure}[p]
    \centering
    \begin{subfigure}{\textwidth}
        \centering
        \vfill
        \includegraphics[width = \textwidth]{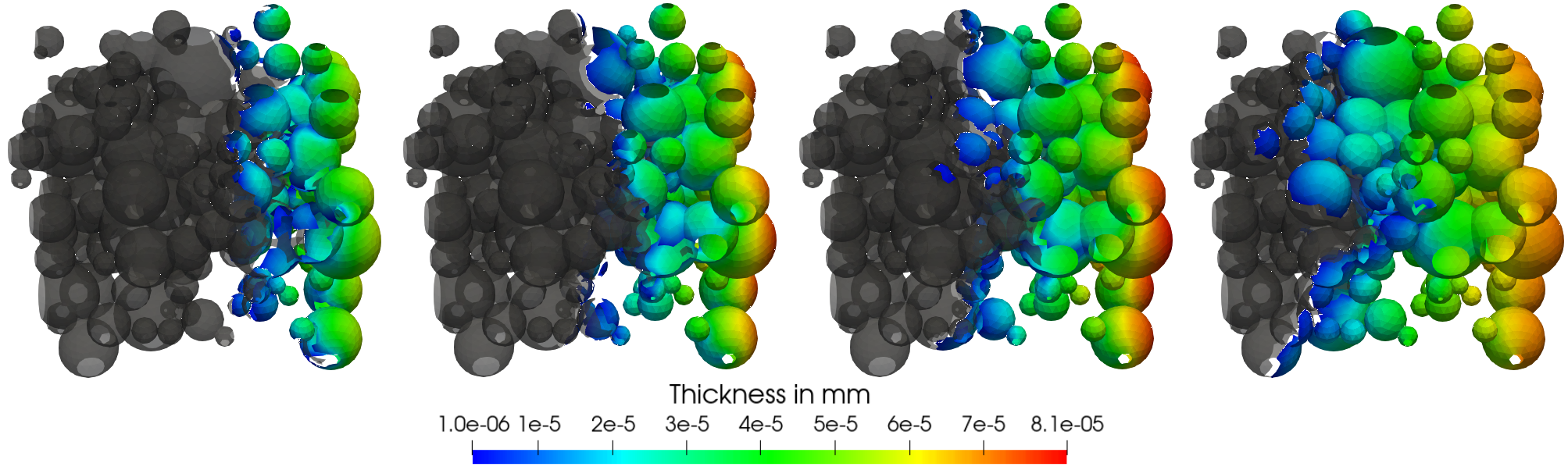}
        \caption{Propagation of the thickness of the \ac{scl} through the geometrically resolved microstructure at the beginning of the formation process at $t=\{0.25, 1, 2, 5\} \ \text{s}$.}
        \label{fig:realistic_thickness_propagation}
        \vfill
        \includegraphics[width = \textwidth]{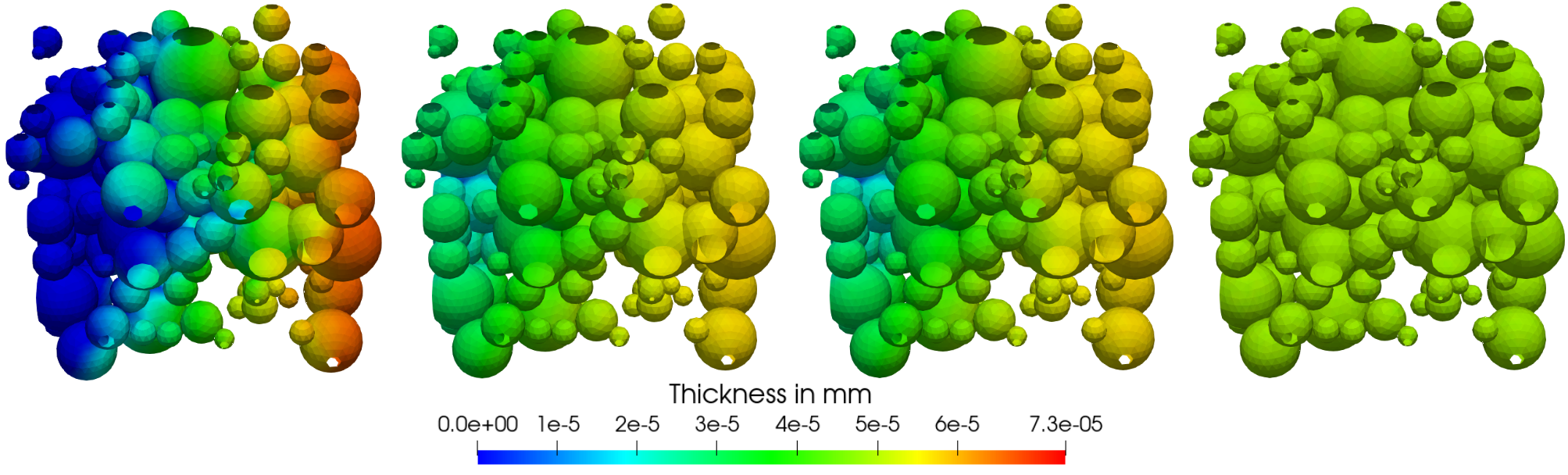}
        \caption{Equalization of the \ac{scl} thickness towards the steady state at $t=\{8.25, 35.25, 65.25, 491.25 \} \ \text{s}$.}
        \label{fig:realistic_thickness_relaxation}
        \vfill
        \definecolor{mycolor1}{rgb}{0.00000,0.44700,0.74100}%
\definecolor{mycolor2}{rgb}{0.85000,0.32500,0.09800}%
\newsavebox{\detailthickness}
\savebox{\detailthickness}{%
\begin{tikzpicture}
\begin{axis}[
width=4.0cm,
height=3.0cm,
xmin=0,
xmax=4,
xtick={0,1,2,3,4},
x tick label style={
/pgf/number format/.cd,
fixed,
fixed zerofill,
precision=1,
/tikz/.cd,
yshift=-.2em},
ymin=0,
ymax=0.081,
ytick={0,0.04,0.08},
y tick label style={
/pgf/number format/.cd,
fixed,
fixed zerofill,
precision=1,
/tikz/.cd},
axis background/.style={fill=white},
axis x line*=bottom,
axis y line*=left,
xmajorgrids,
ymajorgrids
]
\addplot [color=mycolor2, line width=1.0pt, forget plot]
  table[row sep=crcr]{%
0	0\\
0.25	0.0579545267556081\\
0.5	0.0666900797087984\\
0.75	0.0725941441734689\\
1	0.0762124889768436\\
1.25	0.0782636650488738\\
1.5	0.0793946890509865\\
1.75	0.0797764485427739\\
2	0.0797366552190085\\
2.25	0.0794277650903434\\
2.5	0.0788863232442611\\
2.75	0.0782016795697606\\
3	0.077491576460709\\
3.25	0.0767792488861922\\
3.5	0.076057659586439\\
3.75	0.0753089616094009\\
4	0.0745580973451692\\
4.25	0.073850482365877\\
4.5	0.0732186284488257\\
4.75	0.0726212826900322\\
};
\end{axis}
\end{tikzpicture}
}
\begin{tikzpicture}
\begin{axis}[%
width=7.5cm,
height=5.0cm,
scale only axis,
xmin=0,
xmax=500,
xtick={0,100,200,300,400,500},
x tick label style={
/pgf/number format/.cd,
scaled x ticks=base 10:-2,
fixed,
fixed zerofill,
precision=1,
/tikz/.cd,
yshift=-.5em},
xlabel style={font=\color{white!15!black}},
xlabel={Time in s},
ymin=0,
ymax=0.1,
ytick={0,0.02,0.04,0.06,0.08,0.1},
y tick label style={
/pgf/number format/.cd,
scaled y ticks=base 10:1,
fixed,
fixed zerofill,
precision=1,
/tikz/.cd},
ylabel style={font=\color{white!15!black}},
ylabel={Thickness of SCL in $\mu \text{m}$},
axis background/.style={fill=white},
axis x line*=bottom,
axis y line*=left,
xmajorgrids,
ymajorgrids,
legend style={at={(0.95,0.3)}, legend cell align=left, align=left, draw=white!15!black}
]
\addplot [color=mycolor1, line width=1.0pt]
  table[row sep=crcr]{%
0	0\\
0.25	0\\
0.5	0\\
0.75	0\\
1	0\\
1.25	0\\
1.5	0\\
1.75	0\\
2	0\\
2.25	0\\
2.5	0\\
2.75	0\\
3	0\\
3.25	0\\
3.5	0\\
3.75	0\\
4	0\\
4.25	0\\
4.5	0\\
4.75	0\\
5	0\\
5.25	0\\
14.25	0.00328861621615507\\
23.25	0.0125594367314367\\
32.25	0.0209054257003692\\
41.25	0.0267432715439164\\
50.25	0.0308408025891299\\
59.25	0.0338641064489973\\
68.25	0.0361663012712195\\
77.25	0.0380007666933574\\
86.25	0.0394667166961354\\
95.25	0.0406692222499022\\
104.25	0.0416695395249765\\
113.25	0.0424983841878573\\
122.25	0.0432167080528537\\
131.25	0.0438188653086162\\
140.25	0.0443496801997777\\
149.25	0.0448058078028375\\
158.25	0.0452028967848163\\
167.25	0.0455489012458241\\
176.25	0.0458547112812808\\
185.25	0.0461246279956174\\
194.25	0.0463657274346435\\
203.25	0.0465839903212266\\
212.25	0.0467743912237385\\
221.25	0.0469452869297173\\
230.25	0.0470993912679149\\
239.25	0.047229246362174\\
248.25	0.0473504798552647\\
257.25	0.047455756530497\\
266.25	0.0475518881738426\\
275.25	0.0476401030635377\\
284.25	0.0477187873470844\\
293.25	0.0477897362701855\\
302.25	0.0478660895643875\\
311.25	0.0479143472071849\\
320.25	0.0479686303866522\\
329.25	0.0480160895090322\\
338.25	0.0480584468032301\\
347.25	0.0481009389859785\\
356.25	0.0481323742011107\\
365.25	0.04815917316004\\
374.25	0.0481858667839077\\
383.25	0.048214707254963\\
392.25	0.0482411141265309\\
401.25	0.0482680499760687\\
410.25	0.0482877386968552\\
419.25	0.0483102338859703\\
428.25	0.0483181793261\\
437.25	0.048337171118157\\
446.25	0.0483518826701536\\
455.25	0.0483653104221974\\
464.25	0.0483774908267634\\
473.25	0.0483885561606121\\
482.25	0.0483986720170698\\
491.25	0.0484078647396085\\
500.25	0.048416260622292\\
509.25	0.0484239025557914\\
518.25	0.0484308584285934\\
527.25	0.0484372232125586\\
536.25	0.0484430268427778\\
545.25	0.0484483030191338\\
554.25	0.0484531198899487\\
563.25	0.0484575169922386\\
572.25	0.0484615334472112\\
581.25	0.0484652014562908\\
590.25	0.0484685564319295\\
599.25	0.0484716015697214\\
608.25	0.0484743750611451\\
617.25	0.0484769345784228\\
626.25	0.0484792632185985\\
635.25	0.048481406517134\\
644.25	0.0484833413862232\\
653.25	0.0484851197413611\\
};
\addlegendentry{$d_\text{SCL,c,min}$}

\addplot [color=mycolor2, line width=1.0pt]
  table[row sep=crcr]{%
0	0\\
0.25	0.0579545267556081\\
0.5	0.0666900797087984\\
0.75	0.0725941441734689\\
1	0.0762124889768436\\
1.25	0.0782636650488738\\
1.5	0.0793946890509865\\
1.75	0.0797764485427739\\
2	0.0797366552190085\\
2.25	0.0794277650903434\\
2.5	0.0788863232442611\\
2.75	0.0782016795697606\\
3	0.077491576460709\\
3.25	0.0767792488861922\\
3.5	0.076057659586439\\
3.75	0.0753089616094009\\
4	0.0745580973451692\\
4.25	0.073850482365877\\
4.5	0.0732186284488257\\
4.75	0.0726212826900322\\
5	0.0713466180991873\\
5.25	0.0713466180991873\\
14.25	0.0623284330610881\\
23.25	0.0594377069175174\\
32.25	0.0577303080802846\\
41.25	0.0565682264060622\\
50.25	0.0556438060298481\\
59.25	0.0548719866404705\\
68.25	0.054237756278153\\
77.25	0.053666951275159\\
86.25	0.053201816235443\\
95.25	0.0527804638405258\\
104.25	0.0523863334577462\\
113.25	0.0520284598556984\\
122.25	0.0517288837399439\\
131.25	0.0514480622939027\\
140.25	0.0511847440627593\\
149.25	0.0509381508644485\\
158.25	0.0507405036986491\\
167.25	0.0505575080754491\\
176.25	0.0503866838259455\\
185.25	0.0502274913560087\\
194.25	0.0500794375308029\\
203.25	0.0499417877444783\\
212.25	0.0498140844114022\\
221.25	0.0496956225137874\\
230.25	0.0495862291805864\\
239.25	0.0494954331382491\\
248.25	0.0494157138028133\\
257.25	0.0493422456341043\\
266.25	0.0492745674909691\\
275.25	0.0492122154749774\\
284.25	0.0491549933644001\\
293.25	0.0491023665774018\\
302.25	0.0490540666164013\\
311.25	0.0490097880007348\\
320.25	0.0489690340043183\\
329.25	0.0489318694007135\\
338.25	0.0488977898077424\\
347.25	0.0488664742365396\\
356.25	0.048838017014252\\
365.25	0.0488141850125655\\
374.25	0.048790179022354\\
383.25	0.0487679809216646\\
392.25	0.0487478980993343\\
401.25	0.0487295574846689\\
410.25	0.0487127344753466\\
419.25	0.0486973397711961\\
428.25	0.0486833228716675\\
437.25	0.0486705263794623\\
446.25	0.0486588653760219\\
455.25	0.0486481479520834\\
464.25	0.048638464797898\\
473.25	0.0486294213668333\\
482.25	0.0486213967754329\\
491.25	0.0486139193576492\\
500.25	0.04860716334323\\
509.25	0.0486010235673978\\
518.25	0.0485952899431778\\
527.25	0.048590187468677\\
536.25	0.0485854392190855\\
545.25	0.0485811962467998\\
554.25	0.0485773115640276\\
563.25	0.0485736552953681\\
572.25	0.0485703673486496\\
581.25	0.0485674380518325\\
590.25	0.0485646630194706\\
599.25	0.0485621442296592\\
608.25	0.0485598981085612\\
617.25	0.04855780900328\\
626.25	0.0485558932007605\\
635.25	0.0485541273248647\\
644.25	0.0485525415355728\\
653.25	0.0485511435445332\\
};
\addlegendentry{$d_\text{SCL,c,max}$}

\draw (axis cs: 350, 0.075)node{\usebox{\detailthickness}};
\end{axis}
\end{tikzpicture}%
        \caption{Temporal development of maximal and minimal thickness of \ac{scl}. The subfigure represents a zoom into the interval $t = [0;4] \ \text{s}$.}
        \label{fig:realistic_min_max_thickness}
    \end{subfigure}
    \caption{Temporal development of the thickness of the \ac{scl} at the cathode.}
\end{figure}

\begin{figure}[ht]
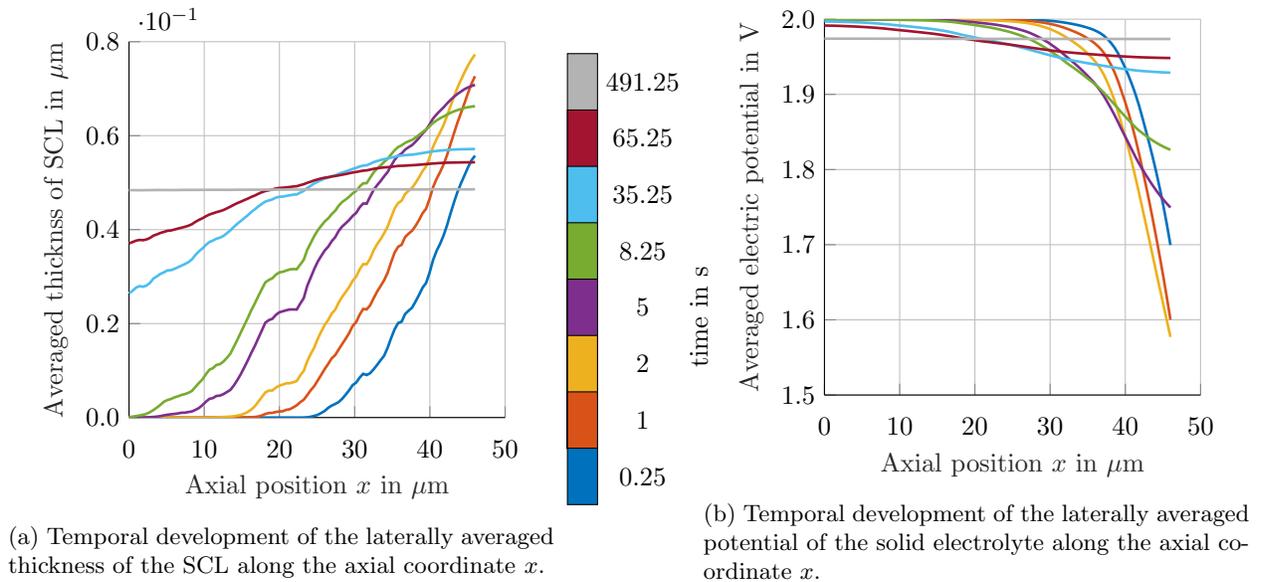

    \centering
    \begin{subfigure}{0.44\textwidth}
        \centering
        \vfill
        \input{figures/realistic_development_thickness_axial.tikz}
        \caption{Temporal development of the laterally averaged thickness of the \ac{scl} along the axial coordinate~$x$.}
        \label{fig:realistic_development_thickness_axial}
        \vfill
    \end{subfigure}
    \hfill
    \begin{subfigure}{0.1\textwidth}
        \centering
        \vfill
        \begin{tikzpicture}

\definecolor{mycolor1}{rgb}{0.00000,0.44700,0.74100}%
\definecolor{mycolor2}{rgb}{0.85000,0.32500,0.09800}%
\definecolor{mycolor3}{rgb}{0.92900,0.69400,0.12500}%
\definecolor{mycolor4}{rgb}{0.49400,0.18400,0.55600}%
\definecolor{mycolor5}{rgb}{0.46600,0.67400,0.18800}%
\definecolor{mycolor6}{rgb}{0.30100,0.74500,0.93300}%
\definecolor{mycolor7}{rgb}{0.63500,0.07800,0.18400}%
\definecolor{mycolor8}{rgb}{0.7 0.7 0.7}%

\filldraw[fill=mycolor8, draw=black] (0,6.0) rectangle (0.4,0.75);
\filldraw[fill=mycolor7, draw=black] (0,5.25) rectangle (0.4,0.75);
\filldraw[fill=mycolor6, draw=black] (0,4.5) rectangle (0.4,0.75);
\filldraw[fill=mycolor5, draw=black] (0,3.75) rectangle (0.4,0.75);
\filldraw[fill=mycolor4, draw=black] (0,3) rectangle (0.4,0.75);
\filldraw[fill=mycolor3, draw=black] (0,2.25) rectangle (0.4,0.75);
\filldraw[fill=mycolor2, draw=black] (0,1.5) rectangle (0.4,0.75);
\filldraw[fill=mycolor1, draw=black] (0,0.0) rectangle (0.4,0.75);

\node at (1.0,0.375) {$0.25$};
\node at (1.0,1.125) {$1$};
\node at (1.0,1.875) {$2$};
\node at (1.0,2.625) {$5$}; 
\node at (1.0,3.375) {$8.25$};
\node at (1.0,4.125) {$35.25$};
\node at (1.0,4.875) {$65.25$};
\node at (1.0,5.625) {$491.25$};

\node at (1.75,2.5) [rotate=90] {time in s};

\end{tikzpicture}
        \vfill
    \end{subfigure}
    \hfill
    \begin{subfigure}{0.44\textwidth}
        \centering
        \vfill
        \input{figures/realistic_development_potential_axial.tikz}
        \caption{Temporal development of the laterally averaged potential of the solid electrolyte along the axial coordinate~$x$.}
        \label{fig:development_potential_axial}
        \vfill      
    \end{subfigure}
    \caption{Laterally averaged quantities.}
\end{figure}

A difference in electric potential between both electrodes is applied. We set the electric potential at the anode to $\Phi_\text{a} = 0 \ \text{V}$ and at the cathode to $\Phi_\text{c} = 2 \ \text{V}$. Again, we want to study the transient behavior until the steady state is reached, such that we choose a total simulation time of $500 \ \text{s}$.\\
At first, we analyze the thickness of the \ac{scl} $d_\text{SCL}$ by plotting it in the three-dimensional geometric representation (see \cref{fig:realistic_thickness_propagation}) during its initial development at $t=\{0.25, 1, 2, 5\} \ \text{s}$.
For visualization, we disable the colorful representation if the thickness is below a threshold $d_\text{SCL}(\vec{x}) < 10^{-6} \ \text{mm}$ and otherwise assign a linear color bar to the thickness. This representation indicates the non-uniformness of the development of the thickness of the \ac{scl}: The dominating trend is comparable to the simplified examples we showed before, namely a propagation through the composite cathode beginning at the points closest to the anode. Due to the geometric complexity of the resolved microstructure, also an inhomogeneous behavior in the lateral plane is observable, which we want to discuss in more detail. The inhomogeneity can be explained by optimal percolation paths. The percolation path in the bulk is now not just defined by the theoretically shortest distance to the anode, but also by geometric obstacles increasing the percolation path, namely the active material particles. Obviously, these obstacles differ in the lateral plane and thus, explain the lateral inhomogeneity of $\Phi_\text{SCL}$. We observe that the \ac{scl} has already further developed where only very few active material particles are on the percolation path.\\
After an \ac{scl} has developed everywhere, a convergence towards an equal-sized thickness is observable. Again, we identify some areas close to the anode where a decrease in thickness occurs, such that the largest thickness is not present in the steady state, but after some instances of time. \cref{fig:realistic_thickness_relaxation} shows the thickness of the \ac{scl} at $t=\{8.25, 35.25, 65.25, 491.25 \} \ \text{s}$.\\
The observed inhomogeneous development of the thickness of the \ac{scl} can further be expressed in terms of maximal and minimal thicknesses $d_\text{SCL,c,min} = \text{min}(d_\text{SCL,c})$ and $d_\text{SCL,c,max} = \text{max}(d_\text{SCL,c})$, respectively (see \cref{fig:realistic_min_max_thickness}). While the maximal value of the thickness is reached within the first instances of time, the \ac{scl} at other positions has not yet developed at all.\\
As the dominating trend of all quantities is one-dimensional, we define laterally averaged quantities as
\begin{equation}
    \bar{\Psi}(x) = \frac{\int_y \int_z \Psi(x,y,z) \dd z \dd y}{\int_y \int_z \dd z \dd y}.
\end{equation}
In \cref{fig:realistic_development_thickness_axial} we show the development of the laterally averaged thickness of the \ac{scl} at the cathode~$\bar{d}_\text{SCL}(x)$ as a function of the axial position $x$.
At $x=x_\text{max}$ the anode is nearest, while at $x=0$ the distance to the anode is greatest. The different lines represent different points in time. Again, we see, the thickness developing through the solid electrolyte and converging towards a constant value at the steady state. Additionally, we observe the influence of the heterogenous geometry by the kink and almost horizontal line at $x \approx 20 \ \mu \text{m}$. We would expect a smooth slope of the averaged thickness if the lateral dimensions would converge towards infinity. However, due to the finite length of the lateral dimension, the lateral planes geometrically differ from each other in a statistical sense and geometric inhomogeneities become visible. This is in good agreement with the observations in \cref{fig:realistic_thickness_propagation}, where an elongated percolation path is visible at the same location where the slope has the prominent kink in \cref{fig:realistic_development_thickness_axial}.\\
Finally, we want to investigate the influence of the \ac{scl} on the bulk domain of the solid electrolyte by analyzing the laterally averaged electric potential $\bar{\Phi}(x)$ and plotting it over the axial direction (see \cref{fig:development_potential_axial}). The lines depict different points in time. At the beginning of the formation of the \acp{scl}, we see that there is only a gradient in electric potential in regions close to $x=x_\text{max}$, such that only there a development of the \ac{scl} is present. During the transient development, the \ac{scl} in the vicinity of $x=x_\text{max}$ is nearly fully charged, such that the optimal percolation paths change towards regions at $x=0$ until the electric potential has converged to a constant value in the steady state. Again, the electric potential is different from the potential computed in the examples before due to different interface areas at the anode and at the cathode side, as already discussed before.
\section*{Conclusions}
We propose a novel approach to incorporate the effect of \acp{scl} spatially resolved into a continuum model for all-solid-state batteries. To our knowledge, this is the first work reporting an approach to model the formation of \acp{scl} in geometrically complex resolved microstructures by overcoming the computational limitations hindering the solution of fully resolved \acp{scl} so far. The governing equations are already established in the literature \cite{Braun2015, BeckerSteinberger2021} and are thermodynamically consistently derived ensuring positive a production rate of entropy. Our approach is motivated by the dominating one-dimensional nature of \acp{scl} as we have shown in this work. We divide the domain of the solid electrolyte into a domain that is close to the electrodes and the remaining domain. Inside the first domain, the partial differential equations are treated as one-dimensional while in the latter, we model them in three dimensions of space. This is advantageous, as a fine discretization, which is required in regions where \acp{scl} develop, is now only required in one dimension of space. This significantly reduces the size of the used mesh and thus, enables a solution of the effects in \acp{scl} even in complex and realistic cases. Furthermore, we assume that the cation concentration in the domain outside of the \ac{scl} region remains constant and subsequently simplify the equations inside this domain. Obviously, the proposed modeling approach introduces approximations into the system. We systematically determine, analyze, and quantify these approximations and show conservation properties of the model. Finally, we present the applicability of our model to realistic microstructures. Beyond the existing knowledge on the temporal development of \acp{scl} in a one-dimensional setup, we observe a strong influence of the geometric inhomogeneity, like non-monotonic development of the thickness of the \ac{scl} or the inhomogeneous convergence of the electric potential towards the steady state.\\
The outlined model can in principle be applied to the case including mass transfer across the interface between electrode and electrolyte. However, a thermodynamically consistent model for the underlying kinetics is not yet known to the authors so far. Geometric variations like incorporating grain boundaries, as well as a systematic variation of the material parameters, are easily viable with the proposed model and should be the subject of future studies.
\section*{Funding}
We gratefully acknowledge support by the Bavarian Ministry of Economic Affairs, Regional Development and Energy [project ``Industrialisierbarkeit von Festk\"orperelektrolytzellen''] and the German Federal Ministry of Education and Research [project FestBatt~2 (03XP0435B)].
\appendixtitleon
\setcounter{table}{0}
\renewcommand{\thetable}{\Alph{section}.\arabic{table}}
\setcounter{equation}{0}
\renewcommand{\theequation}{\Alph{section}.\arabic{equation}}
\setcounter{figure}{0}
\renewcommand{\thefigure}{\Alph{section}.\arabic{figure}}
\begin{appendices}
\section{Summary of a thermodynamically consistent model for solid electrolytes including \acp{scl}}
\label{sec:derivation_scl_model}
We recall an approach to model solid electrolytes including \acp{scl} as already derived in~\cite{Braun2015, BeckerSteinberger2021}, and define a consistent set of boundary and initial conditions.
\subsection*{Governing equations}
We will only summarize the underlying assumptions for the model of \acp{scl} and summarize the resulting equations. For a more thorough derivation, we refer to recent work~\cite{BeckerSteinberger2021}.\\
For the derivations of the set of equations, the continuum approach is followed. Only one species of cations (subscript~$+$) is assumed as a mobile charge carrier within a stationary anion (subscript~$-$) lattice. This corresponds to a transference number~$t_+$ of one and is a valid assumption for various commonly used solid electrolyte materials~\cite{Swift2021}. We refer to the cation concentration as~$c_+$, with~$c_+ \in \ ]0,c_{\text{+,max}}[$. The motion of the cations is expressed by the flux vector~$\vec{N}_{+}$. Thus, the conservation of cations is ensured by
\begin{equation}
    \label{eq:conservation_of_cations}
    \parder{c_+}{t} + \nabla \cdot \vec{N}_{+} = 0 .    
\end{equation}
The free charge $q_\text{F}$ reflects the sum of all charged species and is calculated as the sum of the charge induced by the anions and the cations
\begin{equation}
    \label{eq:free_charge}
    q_\text{F} = \sum_i z_i F c_i ,
\end{equation}
with $i \in \{+,-\}$, $F$ denoting the Faraday constant, and~$z_i$ representing the charge number of species~$i$.\\
A spatially constant dielectric susceptibility~$\chi$ can be assigned to the polarizable background lattice by neglecting any polarization of the cations. The local polarization density $\vec{P}$ is therefore given by $\vec{P} = \epsilon_0 \chi \vec{E}$. Hence, the electric potential~$\Phi$ can be calculated depending on the free charge~$q_\text{F}$ and the dielectric permeability~$\epsilon = \epsilon_0 (1+\chi)$ as
\begin{equation}
    \label{eq:coulombs_equation}
    -\nabla \cdot (\epsilon \nabla \Phi) = q_{\text F} .
\end{equation}
Furthermore, the conservation of charge is decoupled from the conservation of mass and represents another independent equation. Both the free charge density $q_\text{F}$ and the bound charge density $q_\text B = -\nabla\cdot\vec P$, which is the source of the local polarization, contribute to the total charge density $q$, i.e.~$q = q_\text{F} + q_\text{B}$. Each of these quantities are conserved, which allows formulating the conservation of $q_\text{F}$ and $q_\text{B}$
\begin{equation}
     \parder{q_{\text{F,B}}}{t} + \nabla \cdot \vec{i}_{\text{F,B}} = 0 . 
     \label{eq:free_bound_charge}
\end{equation}
The charge transfer inside the solid electrolyte is not just caused by a free current $\vec{i}_\text{F}$ due to the redistribution of cations. Likewise, a polarization current $\vec{i}_\text{B}$ related to the transport of $q_\text{B}$ contributes to the total current density $\vec{i} = \vec{i}_\text{F} + \vec{i}_\text{B}$. Thereby, the current of free charge is derived as~$\vec{i}_\text{F} = \sum_i z_i F \vec{N}_i$, with~$\vec{N}_- = 0$ due to the fixed anion lattice. From~\cref{eq:free_bound_charge} and the definition of $q_\text{B}$ it follows that~$\vec{i}_\text{B} = \parder{\vec{P}}{t} = \parder{\left(\epsilon_0 \chi \vec{E} \right)}{t}$, with~$\vec{E} = - \nabla \Phi$. Finally, the conservation of the total charge $\parder{q}{t} + \nabla \cdot \vec{i} = 0$ reads
\begin{equation}
    \label{eq:conservation_of_charge}
    \parder{q}{t} + \nabla \cdot \left (z_+ F \vec{N}_+ -\epsilon_0 \chi \parder{\nabla \Phi}{t} \right ) = 0 .    
\end{equation}
Additionally, the conservation of linear momentum has to be satisfied. Changes of the linear momentum are caused by pressure gradients $\nabla p$ and electromagnetic forces~$q_\text{F} \vec{E}$ and could be incorporated into a coupled electro-chemo-mechanics model for all-solid-state batteries as introduced in~\cite{Schmidt2022}. However, time-scale considerations~\cite{Braun2015} suggest that inertial forces have a negligible impact compared to the remaining forces, which allows stating the conservation of linear momentum as 
\begin{equation}
    \label{eq:conservation_of_momentum}
    \nabla p = - q_{\text F} \nabla \Phi,
\end{equation}
such that the mechanical pressure~$p$ can be post-processed. In order to close the system of equations, a constitutive law for the flux of cations,~$\vec{N}_+$ has to be specified. An approach based on the free energy guarantees a positive entropy production rate~$\dot{s}_{\text{gen}} (\vec{x},t) > 0$ to derive a linear relation between the gradients $\nabla c_+$, $\nabla \Phi$, and~$\nabla p$ and the flux of cations, $\vec{N}_{+}$. By making use of the relation between~$\nabla p$ and~$\nabla \Phi$ in \cref{eq:conservation_of_momentum}, $\vec{N}_{+}$ can be formulated, such that~$\vec{N}_{+}$ solely depends on~$\nabla c_+$ and~$\nabla \Phi$, the respective diffusion coefficient~$D_+$, and the ionic conductivity~$\sigma$. Both material parameters are a function of the cation concentration and the mobility factor~$\mathcal{L_{++}}$ (see~\cite{Braun2015})
\begin{align}
    \sigma &= (z_+ F)^2 \mathcal{L_{++}} (1-(c_{\text{max}} - c_+) c_+ \Delta \nu) , \\
    D_+ &= \mathcal{L_{++}} R T \frac{c_{\text{max}}}{(c_{\text{max}} -c_+) c_+} ,
\end{align}
with~$\Delta \nu$ the difference in partial molar volumes of cations and cation sites, $T$ the temperature, and $R$ the universal gas constant. For the evaluation of the diffusion coefficient, we assume $c_+ = c_{+,\epsilon}$, if $c_+ < c_{+,\epsilon}$, and $c_+ = c_\text{max} - c_{+,\epsilon}$, if $c_+ > c_\text{max} - c_{+,\epsilon}$ with small values for~$c_{+,\epsilon}$ to avoid divisions by zero during the nonlinear solution scheme. Finally, we summarize the system of equations
\begin{alignat}{2}
\parder{c_+}{t} + \nabla \cdot \vec{N}_{+} &= 0 &&\text{in} \ \Omega_\text{SE}, \\
\parder{q}{t} + \nabla \cdot \left (z_+ F \vec{N}_{+} -\epsilon_0 \chi \parder{\nabla \Phi}{t}\right ) &= 0 &&\text{in} \ \Omega_\text{SE}, \\
-\nabla \cdot (\epsilon \nabla \Phi) &= q_{\text F}  &&\text{in} \ \Omega_\text{SE}, \\
\vec{N}_{+}  &= - D_+ \nabla c_+ - \frac{\sigma}{z_+ F} \nabla \Phi  \quad &&\text{in} \ \Omega_\text{SE}.
\end{alignat}
\end{appendices}
%
\bibliographystyle{IEEEtran}
\bibliography{literature}
\end{document}